\documentclass[review]{elsarticle}

\usepackage{hyperref}

\journal{Journal of Computational Physics}

\usepackage{amsmath,amssymb,bm}
\usepackage{xcolor}
\usepackage{datetime}
\usepackage{subcaption}
\usepackage{setspace}
\usepackage[ruled,vlined,english]{algorithm2e}
\usepackage[babel=true]{csquotes}

\definecolor{Reviewer1}{rgb}{0,0,0}
\definecolor{Reviewer2}{rgb}{0,0,0}
\usepackage{mathrsfs}
\usepackage{amsbsy}

\begin{document}

\begin{frontmatter}

\title{\textcolor{Reviewer1}{Optimized parametric inference for the inner loop of the Multigrid Ensemble Kalman Filter}}

\author[Poitiers]{G. Moldovan\corref{correspondingauthor}}
\cortext[correspondingauthor]{Corresponding author, \textit{gabriel-ionut.moldovan@ensma.fr}}
\author[Poitiers]{G. Lehnasch}
\author[Poitiers]{L. Cordier}
\author[Poitiers]{M. Meldi}
\address[Poitiers]{Institut Pprime, CNRS -
ISAE-ENSMA - Universit\'{e} de Poitiers, 11 Bd. Marie et Pierre Curie,
Site du Futuroscope, TSA 41123, 86073 Poitiers Cedex 9, France}

\begin{abstract}

Essential features of the Multigrid Ensemble Kalman Filter (G. Moldovan, G. Lehnasch, L. Cordier, M. Meldi, \textit{A multigrid/ensemble Kalman filter strategy for assimilation of unsteady flows}, Journal of Computational Physics $443-110481$) recently proposed for Data Assimilation of fluid flows are investigated and assessed in this article. The analysis is focused on the improvement in performance due to the \textit{inner} loop. In this step, data from solutions calculated on the higher resolution levels of the multigrid approach are used as surrogate observations to improve the model prediction on the coarsest levels of the grid. The latter represents the level of resolution used to run the ensemble members for global Data Assimilation.
The method is tested over two classical one-dimensional problems, namely the linear advection problem and the Burgers' equation. The analyses encompass a number of different aspects such as different grid resolutions. The results indicate that the contribution of the \textit{inner} loop is essential in obtaining accurate flow reconstruction and global parametric optimization. These findings open exciting perspectives of application to grid-dependent reduced-order models extensively used in fluid mechanics applications for complex flows, such as Large Eddy Simulation (LES). 

\end{abstract}


\end{frontmatter}



\section{Introduction}
\label{sec:introduction}

The analysis and control of complex configurations for high Reynolds problems of industrial interest is one of the most distinctive open challenges that the scientific community has to face for fluid mechanics applications in the coming decades. The essential non-linearity of this class of flows is responsible for multiscale interactions and an extreme sensitivity to minimal variations in the set-up of the problem. Under this perspective, applications using data-driven tools from Data Assimilation \cite{Daley1991_cambridge,Simon2006_wiley}, and in particular of sequential tools such as the Kalman filter (KF) \cite{Kalman1960_jbe} or the ensemble Kalman filter (EnKF) \cite{Evensen2009_Springer,Asch2016_SIAM}, have been recently used to obtain a precise estimation of the physical flow state accounting for bias or uncertainty in the performance of the investigative tool \cite{Rochoux2014_nhess,Kato2015_jcp,Meldi2017_jcp,Meldi2018_ftc,Labahn2019_pci,Zhang2020_cf,Zhang2020_jcp}. Advances in EnKF Data Assimilation for meteorological application have inspired early studies in computational fluid dynamics (CFD), dealing in particular with the statistical inference of boundary conditions \cite{Mons2016_jcp} or the optimization of the behavior of turbulence / subgrid scale modeling \cite{Wilcox1988_AIAA,Pope2000_cambridge,Durbin2001_Wiley}. However, further advances are needed for a systematic application to complex flows. The first problematic aspect is associated with the computational cost associated with the generation of the ensemble. This issue has usually been bypassed relying on the use of stationary reduced-order numerical simulations such as RANS \cite{Xiao2019_pas}, providing some statistical inference of turbulence modeling \cite{Kato2015_jcp,Xiao2016_jcp,Zhang2020_cf,Zhang2020_jcp}. Applications for scale resolving unstationary flows such as direct numerical simulation (DNS) and large eddy simulation (LES) are much more rare in the literature, because of the computational resources required \cite{Labahn2019_pci,Mons2019_jcp} to generate an acceptably large database to perform a converged parametric inference. Alternative strategies recently explored to reduce the computational costs deal with multilevel \cite{Hoel2016_SIAM,Siripatana2019_cg,Fossum2020_cg} / multifidelity \cite{Popov2021_SIAM} ensemble applications.

A second issue to be faced is that the solution at the end of the time step should, as much as possible, comply with the dynamic equations represented by the discretized numerical model (\emph{conservativity}). One may argue that, if the solution at the beginning of the time step is not accurate, then conservativity is not an efficient objective to be targeted. However, it is also true that the correction performed to the state estimation by KF methods may be responsible for non-physical perturbations of the predicted state, which may produce irreversible numerical instabilities for complex flows.      

In a recent work, the Authors have presented a multigrid ensemble Kalman filter \cite{Moldovan2021_jcp} (MEnKF, renamed from now on MGEnKF). This strategy manipulates data using multiple meshes with different resolutions, exploiting the natural multilevel structure of multigrid solvers. 
\textcolor{Reviewer1}{This algorithm belongs to the class of multilevel methods whose modern treatment was first proposed by Giles \cite{giles_2008, giles_2015}.} The main advantage of the MGEnKF is that a good level of accuracy of the DA procedure (comparable to classical application of the EnKF) is obtained with a significant reduction of the computational resources required. In addition, spurious oscillations of the physical variables due to the state estimation are naturally damped by the iterative resolution procedure of the multigrid solver. In the case of the classical \textcolor{Reviewer2}{Full Approximation Scheme} (FAS)  two-grid multigrid algorithm, the sources of information operating in the MGEnKF are the following:

\begin{itemize}
\item \emph{One main simulation} whose final solution at each time step is provided on the fine level of the grid. 
\item \emph{An ensemble of low-resolution simulations}, which are performed at the coarse level of the grid. 
\item Some \emph{observations} which are provided locally in space and time in the physical domain.
\end{itemize}

The data assimilation procedure, which will be described in detail in Sec.~\ref{sec:maths}, relies on the recursive nature of the multigrid algorithm. The update of the physical state of the flow as well as its parametric description are obtained via two sets of operations:

\begin{itemize}
\item In the \emph{outer loop}, a classical EnKF procedure is performed using the results from an ensemble of simulations and the observation. The EnKF is used to update the physical state and the parametric description of \textit{i)} every member of the ensemble on the coarse grid and \textit{ii)} the main simulation on the fine grid level, via a projection operator.
\item In the \emph{inner loop}, the physical state obtained with the main simulation, which is more precise than the predicted state by the ensemble members, is used as surrogate observation in a new optimization procedure to improve the predictive capabilities of the physical model over the coarse grid. 
\end{itemize} 

In the first article detailing the MGEnKF \cite{Moldovan2021_jcp}, the focus of the analysis has been over the performance of the \emph{outer loop}. This choice was performed due to the central contribution of this loop to the global data assimilation strategy. For this reason, the \emph{inner loop} was suppressed in order to obtain an unambiguous assessment of this element of the MGEnKF.

In this manuscript, an extensive analysis is performed to assess the effects of the \emph{inner loop} over the global accuracy obtained via the MGEnKF. While the accuracy of the numerical model employed to obtain the predicted states for the ensemble members is directly affected by the \emph{outer loop}, further significant improvement is expected with the application of the \emph{inner loop} for two main reasons. The first one is that the usage of surrogate observation from the main simulation is \emph{consistent} with the numerical model used for time advancement on the different refinement levels of the computational grids. Therefore, biases that can affect data assimilation using very different sources of information (such as experiments and numerical results) are naturally excluded. One can also expect a faster rate of convergence owing to this property. The second valuable feature of the \emph{inner loop} is that, as the whole physical state of the main simulation is known on the fine grid level, the surrogate observation can be sampled everywhere in the physical domain. One of the main problematic aspects when assimilating experimental results in numerical models is that often the placement of sensors is affected by physical limitations which can preclude the sampling in highly sensitive locations. This problem is completely bypassed in the \emph{inner loop}, where the user can arbitrarily select the number and the location of sensors. This also opens perspectives of automatic procedures to determine optimal sensor placement \cite{Mons2017_jfm,Mons2021_jfm}, which are not explored in the present manuscript but will be targeted in future works.

The manuscript is organized as follows. In Sec.~\ref{sec:maths}, the mathematical and numerical ingredients used in the framework of the MGEnKF model are introduced and discussed. An extensive discussion of the strategy used to perform the \emph{inner loop} is provided. In Sec.~\ref{sec:apriorianalysis}, the numerical models optimized in the \emph{inner loop} are discussed. In Sec.~\ref{sec:advection}, the results of the numerical simulations for the analysis of the one-dimensional advection equation are provided and discussed. In Sec.~\ref{sec:Burgers}, the analysis is extended to the one-dimensional Burgers' equation. In this case, the accurate representation of non-linear effects is investigated. For both test cases, comparisons between results obtained via the MGEnKF with or without \emph{inner} loop are performed. In Sec.~\ref{sec:conclusions}, concluding remarks and future developments are discussed. 

\section{Ensemble Kalman Filter (EnKF) and multigrid version (MGEnKF)}
\label{sec:maths}
\subsection{Ensemble Kalman Filter (EnKF)}
\label{sec:EnKF_Description}
The Kalman Filter (KF) \cite{Kalman1960_jbe} is a sequential data assimilation tool which provides an estimation of a state variable at time $k$ ($\mathbf{x}_k$), combining the initial estimate $\mathbf{x}_0$, a set of observations $\mathbf{y}^\text{o}_k$, and a linear dynamical model $\mathbf{M}_{k:k-1}$. The state estimation is obtained via two successive operations, which are usually referred to as \emph{forecast} step (superscript $f$) and \emph{analysis} step (superscript $a$). The classical version of the algorithm reads as follows:

\begin{subequations}
\begin{align}
\mathbf{x}^\text{f}_{k} & = \mathbf{M}_{k:k-1} \mathbf{x}^\text{a}_{k-1}\label{eq:x_mdl_kf}\\
\mathbf{P}^\text{f}_{k}& = \mathbf{M}_{k:k-1} \mathbf{P}^\text{a}_{k-1} {\mathbf{M}^\top_{k:k-1}}+\mathbf{Q}_k \label{eq:P_mdl_kf} \\
  \mathbf{K}_k& = \mathbf{P}^\text{f}_{k}{\mathbf{H}^\top_k}\left(\mathbf{H}_k \mathbf{P}^\text{f}_{k} {\mathbf{H}^\top_k}+\mathbf{R}_k\right)^{-1}\label{eq:k_kf}\\
  \mathbf{x}^\text{a}_{k}& = \mathbf{x}^\text{f}_{k} + \mathbf{K}_k\left(\mathbf{y}^\text{o}_k-\mathbf{H}_k\mathbf{x}^\text{f}_{k}\right)\label{eq:x_kf}\\
  \mathbf{P}^\text{a}_{k}& = \left(I-\mathbf{K}_k \mathbf{H}_k\right)\mathbf{P}^\text{f}_{k}\label{eq:P_kf}
\end{align}
\end{subequations}

The final state estimation $\mathbf{x}^\text{a}_{k}$ is obtained in \eqref{eq:x_kf} by combining the predictor solution $\mathbf{x}^\text{f}_{k}$ and a correction term accounting for the real observation $\mathbf{y}^\text{o}_k$ and the observation predicted from $\mathbf{x}^\text{f}_{k}$ through the linear observation operator $\mathbf{H}_k$. This correction term is weighted by the matrix $\mathbf{K}_k$, usually referred to as \emph{Kalman gain}. The matrices $\mathbf{Q}_k$ and $\mathbf{R}_k$ measure the level of confidence in the model and in the observation, respectively. The error covariance matrices $\mathbf{P}^{\text{f}/\text{a}}_k$, defined as $
\mathbb{E}\left[
\left(\mathbf{x}_k^{\text{f}/\text{a}} - \mathbb{E}(\mathbf{x}_k^{\text{f}/\text{a}})\right) 
\left(\mathbf{x}_k^{\text{f}/\text{a}} - \mathbb{E}(\mathbf{x}_k^{\text{f}/\text{a}})\right)^\top \right]$, are advanced in time in the \emph{forecast} step \eqref{eq:P_mdl_kf} and manipulated in the \emph{analysis} step \eqref{eq:P_kf}.    

These last two operations are usually extremely expensive and they may require computational resources that are orders of magnitude larger than the model operation. In order to alleviate such computational issues, one possible alternative is to provide an approximation of $\mathbf{P}$ using stochastic Monte-Carlo sampling. The Ensemble Kalman Filter (EnKF) \cite{Evensen1994,Evensen2009_Springer} provides an approximation of $\mathbf{P}_k$ by means of an ensemble of $N_\text{e}$ states. \textcolor{Reviewer1}{One clear advantage of the EnKF over low-rank approximation of the Kalman filter is that it can naturally advance the state in time using non-linear models.}

Given an ensemble of forecast/analysis states at a given instant $k$, the ensemble matrices collecting the information of the database are defined as:
\begin{equation}
\pmb{\mathscr{E}}_k^{\text{f}/\text{a}}=
\left[\mathbf{x}_k^{\text{f}/\text{a},(1)},\cdots,\mathbf{x}_k^{\text{f}/\text{a},(N_\text{e})}\right]\in\mathbb{R}^{N_x\times N_\text{e}}, \label{eq:Ensemble_Matrix}
\end{equation}

where $N_x$ is the size of the state vector. Starting from the data assembled in $\pmb{\mathscr{E}}_k^{\text{f}/\text{a}}$, the ensemble means 
\begin{equation}
    \overline{\mathbf{x}_k^{\text{f}/\text{a}}}=\frac{1}{N_\text{e}}\sum^{N_\text{e}}_{i=1}\mathbf{x}_k^{\text{f}/\text{a},(i)} \label{eq:Ensemble_Mean}
\end{equation} 

are used to obtain the normalized ensemble anomaly matrices:
\begin{equation}
\mathbf{X}_k^{\text{f}/\text{a}}=\frac{\left[\mathbf{x}_k^{\text{f}/\text{a},(1)}-\overline{\mathbf{x}_k^{\text{f}/\text{a}}},\cdots,\mathbf{x}_k^{\text{f}/\text{a},(N_\text{e})}-\overline{\mathbf{x}_k^{\text{f}/\text{a}}}\right]}{\sqrt{N_\text{e}-1}}\in\mathbb{R}^{N_x\times N_\text{e}}, \label{eq:Ensemble_Anomaly}
\end{equation}

The approximated error covariance matrices, hereafter denoted with the superscript $e$,  are obtained by the matrix products: 
\begin{equation}
\mathbf{P}_k^{\text{f}/\text{a},\text{e}}=\mathbf{X}_k^{\text{f}/\text{a}} \left(\mathbf{X}_k^{\text{f}/\text{a}}\right)^\top\in\mathbb{R}^{N_x\times N_x} \label{eq:Ensemble_P}
\end{equation}

The goal of the EnKF is to mimic the BLUE (Best Linear Unbiased Estimator) analysis of the Kalman filter.
For this,  Burgers et al. \cite{Burgers1998} showed that the observations must be considered as random variables with an average corresponding to the observed values and a covariance $\mathbf{R}_k$ (the so-called \emph{data randomization} trick).
Therefore, given the discrete observation vector $\mathbf{y}^\text{o}_k\in \mathbb{R}^{N_y}$ at an instant $k$, the ensemble of perturbed observations is defined as:
\begin{equation}
\mathbf{y}_k^{\text{o},(i)}=\mathbf{y}_k^\text{o}+\mathbf{\epsilon}_k^{\text{o},(i)},
\quad
\text{with}
\quad
i=1,\cdots,N_\text{e}
\quad
\text{and}
\quad
\mathbf{\epsilon}_k^{\text{o},(i)}\sim \mathcal{N}(0,\mathbf{R}_k)
.\label{eq:perturbed_y}
\end{equation}

A normalized anomaly matrix of the observations errors is defined as
\begin{equation}
\mathbf{E}_k^{\text{o}}=
\frac{1}{\sqrt{N_\text{e}-1}}
\left[
\mathbf{\epsilon}_k^{\text{o},(1)}-\overline{\mathbf{\epsilon}_{k}^{\text{o}}},
\mathbf{\epsilon}_k^{\text{o},(2)}-\overline{\mathbf{\epsilon}_{k}^{\text{o}}},
\cdots,
\mathbf{\epsilon}_k^{\text{o},(N_\text{e})}-\overline{\mathbf{\epsilon}_{k}^{\text{o}}},
\right]
\in\mathbb{R}^{N_y\times N_\text{e}},
\end{equation} 
where
$\displaystyle\overline{\mathbf{\epsilon}_{k}^{\text{o}}}=\frac{1}{N_\text{e}}\sum_{i=1}^{N_\text{e}}\mathbf{\epsilon}_k^{\text{o},(i)}$.

The ensemble covariance matrix of the measurement error can then be estimated as
\begin{equation}
\mathbf{R}_k^\text{e}=\mathbf{E}^\text{o}_k \left(\mathbf{E}^\text{o}_k\right)^\top\in\mathbb{R}^{N_y\times N_y}. \label{eq:observation_matrix_errorcov}
\end{equation}

Combining the previous results, one obtains (see \cite{Asch2016_SIAM}) the standard stochastic EnKF algorithm. The corresponding analysis step consists of updates  performed on each of the ensemble members, as given by 
\begin{equation}
\mathbf{x}_k^{\text{a},(i)}=
\mathbf{x}_k^{\text{f},(i)}+
\mathbf{K}_k^\text{e}
\left(y_k^{\text{o},(i)}-\mathbf{\mathcal{H}}_k\left(\mathbf{x}^{\text{f},(i)}_k\right)\right)\label{eq:ensemble_update}
\end{equation}
where $\mathbf{\mathcal{H}}_k$ is the non-linear observation operator. The expression of the Kalman gain is
\begin{equation}
\mathbf{K}_k^\text{e}=
\mathbf{X}_k^\text{f}
\left(\mathbf{Y}_k^\text{f}\right)^\top
\left(
\mathbf{Y}_k^\text{f}
\left(\mathbf{Y}_k^\text{f}\right)^\top
+
\mathbf{E}_k^\text{o} \left(\mathbf{E}_k^\text{o}\right)^\top
\right)^{-1}
\end{equation}
where $\mathbf{Y}_k^\text{f}=\mathbf{H}_k\mathbf{X}_k^\text{f}$.


Applications of EnKF strategies in fluid mechanics and engineering deal with a number of different topics such as wildfire propagation \cite{Rochoux2014_nhess}, combustion \cite{Labahn2019_pci}, turbulence modeling \cite{Xiao2016_jcp,Zhang2020_cf,Zhang2021_cf} and hybrid variational-EnKF methods \cite{Mons2019_jcp}. 
The state estimation procedure can be extended to infer the values of free parameters $\theta$ which describe the model. Several proposals have been reported in the literature (see the review work by Asch et al. \cite{Asch2016_SIAM} for an extended discussion). The strategy selected for this work is the \emph{dual estimation} proposed by Moradkhani et al. \cite{DENKF_MORADKHANI2005}. 

\subsection{Multigrid Ensemble Kalman Filter (MGEnKF)}
\label{sec:MGEnKF_Description}
While the ensemble approximation $\mathbf{P}^\text{e}$ of the error covariance matrix $\mathbf{P}$ provides a significant reduction of the computational resources required by the Kalman filter, the generation of a database of $40-100$ elements may still be out of reach for complex applications in fluid mechanics. For this reason, the team has developed a multigrid ensemble Kalman filter (MGEnKF) \cite{Moldovan2021_jcp} strategy which exploits the numerical features of multigrid solvers. The algorithm is flexible and can be applied to numerical solvers using multiple refinement levels. Here, for sake of simplicity, it is illustrated on the classical two-grid FAS (Full Approximation Scheme) \cite{Brandt1977_mc,Wesseling1999_jcam} algorithm. This scheme employs a fine grid (solutions indicated with the superscript F) and a coarse grid (superscript C) in a three-step procedure: 

\begin{enumerate}
\item A time advancement is performed starting from the solution $\mathbf{x}^\text{\tiny F}_{k-1}$ to obtain the state $(\mathbf{x}^\text{\tiny F}_{k})^{*}$.
\item $(\mathbf{x}^\text{\tiny F}_{k})^{*}$ is projected on the coarse grid level leading to the state $\left(\mathbf{x}^\text{\tiny C}_k\right)^{*}=
\Pi_\text{\tiny C}\left(\left(\mathbf{x}_k^\text{\tiny F}\right)^{*}\right)
$. Iterative calculations are used to filter out high-frequency noise and derive the solution $\left(\mathbf{x}^\text{\tiny C}_k\right)^{'}$.
\item \textcolor{Reviewer2}{The difference between $\left(\mathbf{x}^\text{\tiny C}_k\right)^{'}$ and $\left(\mathbf{x}^\text{\tiny C}_k\right)^{*}$ is projected on the fine grid leading to the state $(\mathbf{x}^\text{\tiny F}_{k})^{'}=(\mathbf{x}^\text{\tiny F}_{k})^{*} + \Pi_\text{\tiny F}\left(\left(\mathbf{x}_k^\text{\tiny C}\right)^{'}-\left(\mathbf{x}_k^\text{\tiny C}\right)^{*}\right)$. Thereafter, $(\mathbf{x}^\text{\tiny F}_{k})^{'}$ is used as initial condition for a last set of iterations on the fine grid, which produces the final state $\mathbf{x}^\text{\tiny F}_{k}$}.
\end{enumerate}

\begin{figure}[htbp]
\includegraphics[width=0.95\textwidth]{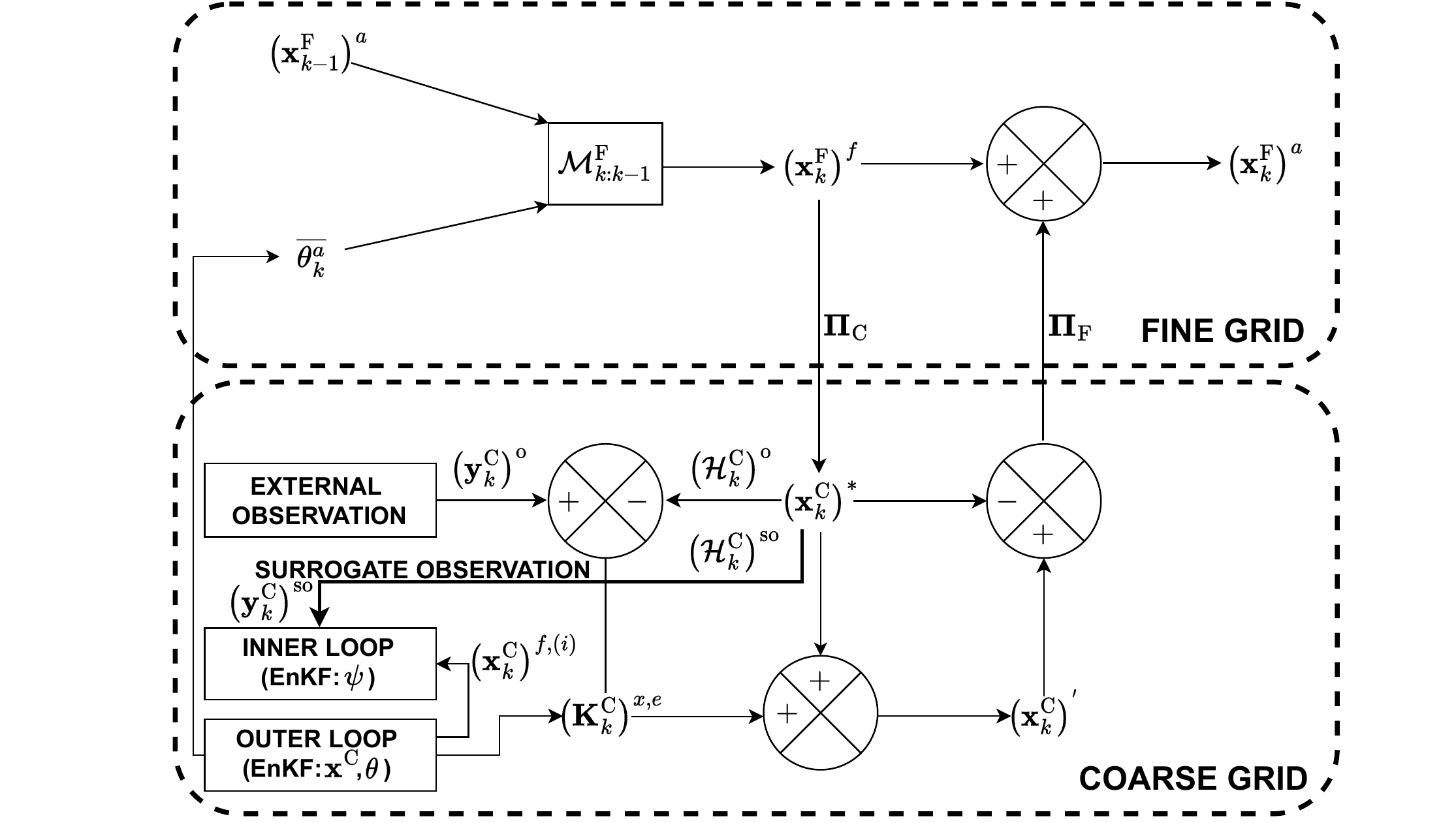}
\caption{\label{fig:schema_MGENKF}
Schematic representation of the Multigrid Ensemble Kalman Filter (MGEnKF) \cite{Moldovan2021_jcp}. Two different levels of representation (fine and coarse grids) are used to obtain an estimation for the main simulation running on the fine grid. The inner and outer loop of the Dual Ensemble Kalman filter procedure are solved using ensemble members generated on the coarse grid.}
\end{figure}

\begin{figure}[htbp]
\includegraphics[width=1\textwidth]{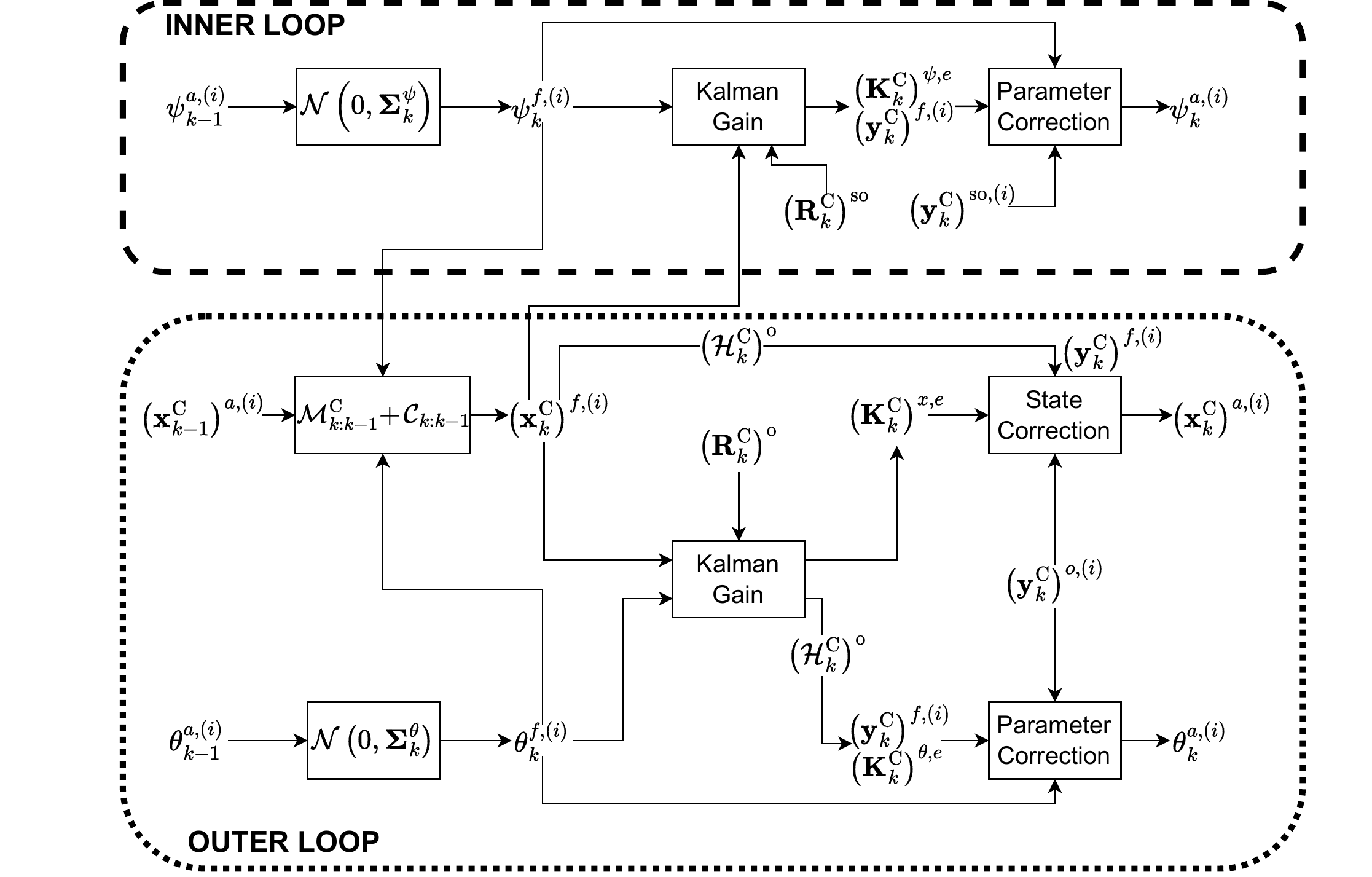}
\caption{\label{fig:schema_inou_loop}
Schematic representation of the boxes \emph{inner} and  \emph{outer loop} in Fig.~\ref{fig:schema_MGENKF}. Two optimization procedures based on EnKF are sequentially performed.}
\end{figure}

A two-level EnKF, shown in Fig.~\ref{fig:schema_MGENKF}, is naturally integrated in the multigrid structure of the algorithm. This reduces the computational costs of the data assimilation procedure and contributes to render the solutions compatible with the dynamic equations of the model \cite{Moldovan2021_jcp}.

\textcolor{Reviewer2}{We emphasize here that in the MGEnKF algorithm, the model used for the time-integration of the state does not necessarily need to use implicit time-discretisation schemes, where multigrid strategies can actually be applied. Depending on the case, explicit time-integration models could also be used. However, this can have a negative impact on the conservativity of the system and it's up to the user to determine the most appropriate strategy for the desired application. Notably, in the 1D experiments performed in this article, the numerical stability of the simulations was not necessarily an issue, thus explicit models were used for simplicity. For 2D and 3D analysis of compressible flows, conservativity can be a real issue, and the use of implicit methods is more robust \cite{Moldovan2021_jcp}. For sake of simplicity and to keep the description general, we have chosen to present the algorithm assuming that implicit methods are used, thus the iterative part of the multigrid method is kept in the description. Whenever we mention classical iterative procedures, we refer to, for instance, the Jacobi method, or the Gauss–Seidel method. The interested reader is referred to Chapter 10 of \cite{Hirsch2007}, which contains a comprehensive review of iterative methods used for solving algebraic systems of equations related to fluid mechanics.}
 
\textcolor{Reviewer2}{The sequential estimation by EnKF is performed in} 
two distinct parts: an \emph{outer loop}, where observations are obtained from an external source of information, and an \emph{inner loop}, where fine grid solutions projected onto the coarse grid are used as surrogate observations \textcolor{Reviewer2}{(see Fig.~\ref{fig:schema_MGENKF})}. 
\textcolor{Reviewer2}{%
Let $\mathbf{\mathcal{M}}^\text{\tiny F}_{k:k-1}\left(
	\mathbf{x}_{k-1}^\text{\tiny F},
	\mathbf{\theta}_{k}\right)$ be a parametrized discretized model defined on the fine grid where $\mathbf{\theta}_{k}$ is a set of free parameters describing the setup of the model.} 
The	complete general algorithm for MGEnKF is structured in the following operations \textcolor{Reviewer2}{(see \ref{sec:DA_algorithms} for the pseudo-algorithms)}:

\begin{enumerate}
\item \textbf{Predictor step}. The analyzed solution defined on the fine grid $\left(\mathbf{x}^\text{\tiny F}_{k-1}\right)^\text{a}$ is used to calculate a forecast state $\left(\mathbf{x}^\text{\tiny F}_{k}\right)^\text{f}$ 
\textcolor{Reviewer2}{
$$
	\left(\mathbf{x}_k^\text{\tiny F}\right)^\text{f}=
	\mathbf{\mathcal{M}}^\text{\tiny F}_{k:k-1}\left(
	\left(\mathbf{x}_{k-1}^\text{\tiny F}\right)^\text{a},
	\overline{\mathbf{\theta}}_{k}^\text{a}\right)
$$
} 
where 
\textcolor{Reviewer2}{$\overline{\mathbf{\theta}}_{k}^\text{a}=\frac{1}{N_\text{e}}\sum_{i=1}^{N_\text{e}}\mathbf{\theta}_{k}^{\text{a},(i)}$ is the ensemble average of coarse-level parameters.} Each member $i$ of the ensemble of states defined on the coarse grid is also advanced in time
$$
	\left(\mathbf{x}_k^\text{\tiny C}\right)^\text{f,(i)}=
	\mathbf{\mathcal{M}}^\text{\tiny C}_{k:k-1}\left(
	\left(\mathbf{x}_{k-1}^\text{\tiny C}\right)^\text{a,(i)},
	\mathbf{\theta}_{k}^\text{f,(i)}\right) + \mathbf{\mathcal{C}}_{k:k-1}\left(
	\left(\mathbf{x}_{k-1}^\text{\tiny C}\right)^\text{a,(i)},
	\mathbf{\psi}_{k}^\text{f,(i)}\right),
$$ 
where $\mathbf{\mathcal{M}}^\text{\tiny C}_{k:k-1}$ is the coarse grid model parameterized by $\mathbf{\theta}_{k}^\text{f,(i)}$, while $\mathbf{\mathcal{C}}_{k:k-1}$ is an additional correction term included to compensate the loss of resolution due to calculations on the coarse grid \cite{Brajard2020_arxiv}. This additional model, whose structure is usually unknown, is driven by the set of free parameters $\mathbf{\psi}_{k}^\text{f,(i)}$ \textcolor{Reviewer2}{(see Sec.~\ref{sec:apriorianalysis} for extended discussion about this correction term)}.
%


\item \textbf{Projection on the coarse grid \& \emph{inner} loop}. $\left(\mathbf{x}^\text{\tiny F}_{k}\right)^\text{f}$ is projected on the coarse grid space via a projection operator $\Pi_\text{\tiny C}$, so that $\left(\mathbf{x}^\text{\tiny C}_{k}\right)^{*}$ is obtained, \textit{i.e.} 
\begin{equation*} \label{eq:CG_projection}
\left(\mathbf{x}^\text{\tiny C}_k\right)^{*}=
\Pi_\text{\tiny C}\left(\left(\mathbf{x}_k^\text{\tiny F}\right)^\text{f}\right).
\end{equation*}
\textcolor{Reviewer2}{
At this step, surrogate observations (denoted hereafter with the superscript \enquote{so}) are determined from $\left(\mathbf{x}^\text{\tiny C}_k\right)^{*}$ with an observation operator $\left(\mathcal{H}_k^\text{\tiny C}\right)^\text{so}$:
\begin{equation*} \label{eq:surrogate_observation}
\left(\mathbf{y}_k^\text{\tiny C}\right)^\text{so}=\left(\mathcal{H}_k^\text{\tiny C}\right)^\text{so}\left(\mathbf{x}^\text{\tiny C}_k\right)^{*}
\end{equation*}
These surrogate observations are used exclusively in the \emph{inner} loop for estimating  the parameters $\mathbf{\psi}_{k}^\text{f,(i)}$.
Hence, the ensemble states $\left(\mathbf{x}_k^\text{\tiny C}\right)^\text{f,(i)}$ are not modified, but the free parameters $\mathbf{\psi}_{k}^\text{f,(i)}$ are updated by a specific \textit{inner loop} EnKF to obtain the analysed values $\mathbf{\psi}_{k}^\text{a,(i)}$. This parameter optimization targets an improvement of the prediction of the ensemble members simulated on the coarse grid via an update of the correction term $\mathbf{\mathcal{C}}_{k:k-1}\left(
	\left(\mathbf{x}_{k-1}^\text{\tiny C}\right)^\text{a,(i)},
	\mathbf{\psi}_{k}^\text{f,(i)}\right)$.
}

\item \textbf{\emph{Outer} loop}. If external observations $\left(\mathbf{y}_k^\text{\tiny C}\right)^\text{o}$ are available, the ensemble forecast $\left(\mathbf{x}^\text{\tiny C}_{k}\right)^{\text{f},(i)}$ is corrected with the standard Dual EnKF procedure to obtain $\left(\mathbf{x}^\text{\tiny C}_{k}\right)^{\text{a},(i)}$ as well as an update of the parameters $\mathbf{\theta}_{k}^{\text{a},(i)}$. 

\item \textbf{Determination of the state variables on the coarse grid}. 
In this step, the coarse grid state is updated. If observations are not available, this update, referred as $\left(\mathbf{x}^\text{\tiny C}_k\right)^{'}$, is obtained by classical iterative procedures on the coarse grid using the initial solution $\left(\mathbf{x}^\text{\tiny C}_{k}\right)^{*}$. 
On the other hand, if observations are available, the Kalman gain matrix $\left(\mathbf{K}_k^\text{\tiny C}\right)^{x,\text{e}}$ determined by EnKF, is used to determine the coarse grid solution $\left(\mathbf{x}^\text{\tiny C}_k\right)^{'}$ through a KF operation, \textit{i.e.} 
		\begin{align*}
\left(\mathbf{x}^\text{\tiny C}_k\right)^{'} & =
\left(\mathbf{x}^\text{\tiny C}_k\right)^{*}+
\left(\mathbf{K}_k^\text{\tiny C}\right)^{x,\text{e}}
\left[
\left(\mathbf{y}_k^\text{\tiny C}\right)^\text{o}-
\left(\mathbf{\mathcal{H}}_k^\text{\tiny C}\right)^\text{o}
\left(\left(\mathbf{x}_k^\text{\tiny C}\right)^{*}\right)
\right]
		\end{align*} 
%
\item \textbf{Estimation on the fine grid}. The fine grid state solution $\left(\mathbf{x}^\text{\tiny F}_k\right)^{'}$ is determined using the results obtained on the coarse space: $\left(\mathbf{x}^\text{\tiny F}_k\right)^{'}=\left(\mathbf{x}^\text{\tiny F}_{k}\right)^\text{f}+\Pi_\text{\tiny F}\left(\left(\mathbf{x}^\text{\tiny C}_k\right)^{'}-\left(\mathbf{x}^\text{\tiny C}_{k}\right)^{*}\right)$. The state $\left(\mathbf{x}^\text{\tiny F}_k\right)^\text{a}$ is obtained from a final iterative procedure starting from $\left(\mathbf{x}^\text{\tiny F}_k\right)^{'}$.
\end{enumerate}

A detailed representation of the \textcolor{Reviewer2}{\emph{inner} and \emph{outer} loops} shown in Fig.~\ref{fig:schema_MGENKF} is provided in Fig.~\ref{fig:schema_inou_loop}. This method is reminiscent of multilevel \cite{Hoel2016_SIAM,Siripatana2019_cg,KodyLaw2020,Fossum2020_cg} and multifidelity \cite{Gorodetsky2020_jcp,Popov2021_SIAM} ensemble techniques for data assimilation. 
\textcolor{Reviewer1}{Our approach differs by the use of surrogate observations issued from the fine grid resolution to infer the correction parameters of the numerical integration scheme. In addition, }
only one main simulation on the fine grid is needed, significantly reducing the computational cost associated to the whole procedure.

\section{Optimization of numerical integration schemes on a coarse mesh}
\label{sec:apriorianalysis}

Numerical schemes unavoidably lead to  diffusive and dispersive errors that can excessively alter the representativity of solutions on too coarse meshes. 
These errors can lead to an imprecise determination of the error covariance matrix, prohibiting the updating of the solution on the fine mesh.
An original feature of the MGEnKF algorithm presented in Sec.~\ref{sec:maths} is that the solution advanced at the most refined level of the grid can also be used as surrogate observation to optimize the parameters of the numerical schemes. 
In this section, we present the principles guiding the choice of the numerical scheme and associated parametrization retained to control the numerical errors on coarse meshes.
Let us consider the prototype 1D linear advection equation of a scalar quantity $u$ advected with the constant velocity $c$:
\begin{equation}
\label{eq:1D_adv}
\frac{\partial u}{\partial t}+c \frac{\partial u}{\partial x}=0
\end{equation}

For simplicity of presentation, we restrict ourselves to using an explicit finite difference scheme on four-point stencils (second or third order accurate) for the case $c>0$.
The spatial discretization is performed on a Cartesian mesh with a constant size $\Delta_x$.
$\Delta_t$ is the time step and  $\sigma=c\Delta_t / \Delta_x$ the CFL number.  $u_\text{j}^\text{k}$ represents the discrete numerical solution at the spatial location $x_\text{j}=(j-1)\Delta_x$ at time $t=k\Delta_t$.
The following considerations can be extended quite easily to non-linear systems with time and space varying advection velocity, irregular mesh or higher-order schemes. 

A general one-parameter family of second order accurate schemes (see \cite{Hirsch2007}, p. 362, Eq.~(8.2.37)) may be defined on a backward upwind stencil $(j-2,j-1,j,j+1)$  as: 
\begin{equation}
\label{eq:schema}
\begin{aligned}
u_\text{j}^\text{k} & = & u_\text{j}^\text{k-1}-\frac{\sigma}{2} \left( u_\text{j+1}^\text{k-1}-u_\text{j-1}^\text{k-1} \right) +
\frac{\sigma^2}{2} \left( u_\text{j+1}^\text{k-1}-2 u_\text{j}^\text{k-1}+u_\text{j-1}^\text{k-1} \right) \\
& & +
\delta \left( -u_\text{j-2}^\text{k-1}+3u_\text{j-1}^\text{k-1}-3u_\text{j}^\text{k-1}+u_\text{j+1}^\text{k-1} \right)
\end{aligned}
\end{equation}

The first line of \eqref{eq:schema} corresponds to the standard centered Lax Wendroff scheme. The second line can be interpreted as the discretization of an additional dispersion term of the form $\delta \Delta_x^3 u_{xxx}$.
The expression \eqref{eq:schema} is compatible with the linear advection equation \eqref{eq:1D_adv} discretized at precision order proportional to $\left(\Delta_x^2, \Delta_t^2\right)$. If one retains the two first dominant error terms in the combination of Taylor expansions corresponding to these discrete terms forming scheme \eqref{eq:schema}, 
the following equivalent differential equation is obtained:
\begin{equation}
\label{eqequiv}
\begin{aligned}
\frac{\partial u}{\partial t}+c \frac{\partial u}{\partial x} & = & \frac{c}{6} \left[ 6\delta -
\sigma \left( 1-\sigma^2\right)\right] \Delta_x^2 \frac{\partial ^3 u}{\partial x^3} & \\
 & & - \frac{c}{8} \left[ \delta \left( 1-2 \sigma \right) + \frac{\sigma^2}{4} \left( 1 - \sigma^2 \right)  \right]  \Delta_x^3 \frac{\partial ^4 u}{\partial x^4} & + \mathcal{O}\left(\Delta_x^4\right)\\
\end{aligned}
\end{equation}
  
Equation \eqref{eqequiv} reveals that the dominant numerical error of \eqref{eq:schema} is dispersive with an error proportional to $\Delta_x^2$ and that a less dominant (third-order) diffusive error term occurs. The expressions of these errors show that their relative level varies both as function of $\sigma$ and $\delta$. 
In order to keep the MGEnKF algorithm sufficiently flexible and general, we will not search to optimize the parameter $\sigma$.
Therefore, it is chosen to target only an optimization for $\delta$  with a strategy suitable for any value of $\sigma$. This choice allows the user to set the value of $\sigma$ based on practical constraints, such as for synchronizing simulations with available observation data.
The level of the dominant error can be controlled through the parameter $\delta$. By setting in particular $\delta = \sigma(1-\sigma^2)/6$, the dominant error term cancels out and the scheme becomes third order accurate in space with now a  diffusive dominant error term proportional to $\Delta_x^3$. Other values of $\delta$ maintain the formal second order accuracy but can induce significantly different effective evolution of the numerical errors. 
\begin{figure}[htbp]
\centering
\includegraphics[width=0.9\textwidth, trim=0cm 0cm 0cm 0cm,clip]{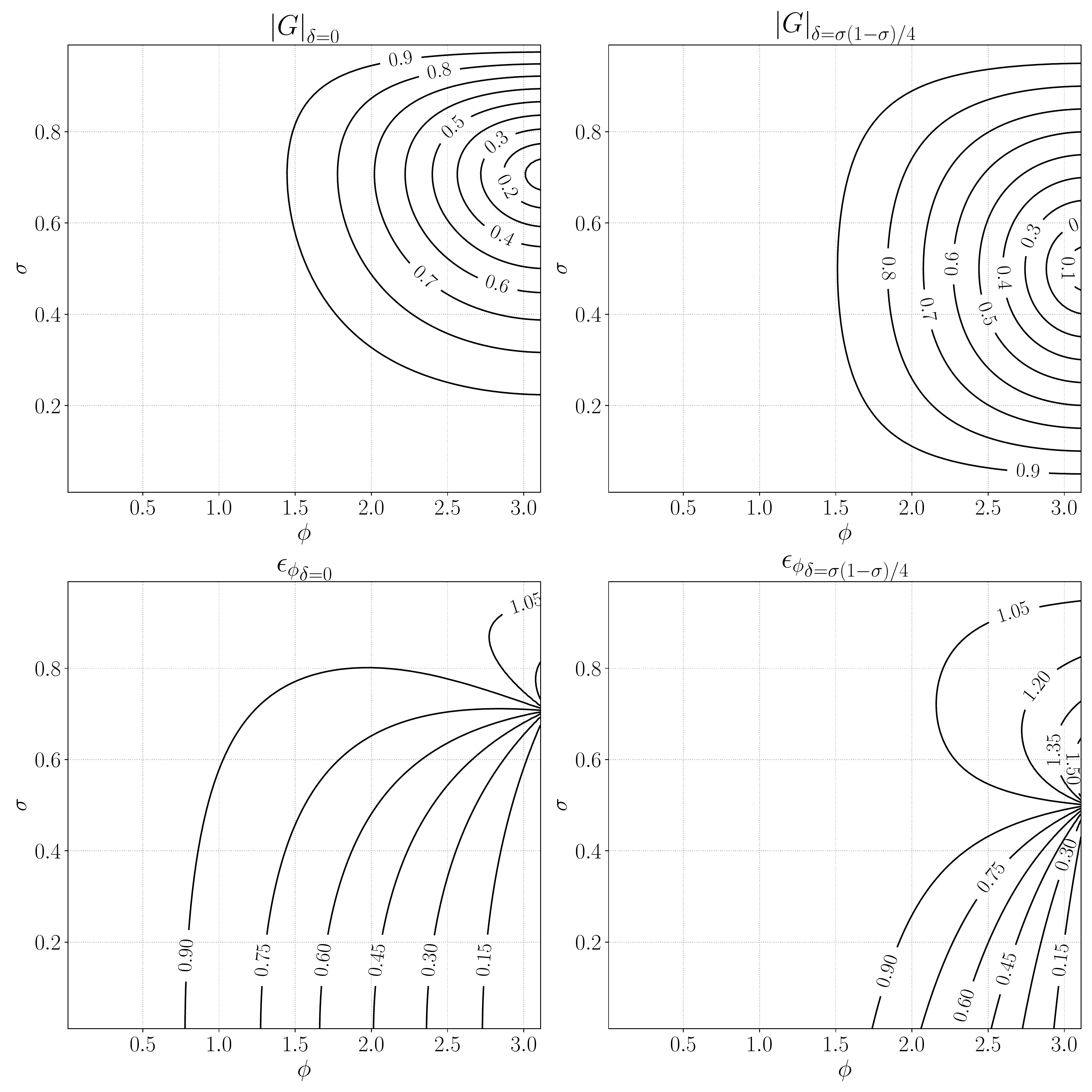}
\caption{\label{fig:figerror} Comparison of diffusive ($\epsilon_D=|G|$, top) and dispersive ($\epsilon_{\phi}$, bottom) errors for two different \textit{a priori} parametrizations of $\delta$. $\delta=0$ (left column) corresponds to the Lax Wendroff scheme and $\delta=\sigma\left(1-\sigma\right)/4$ (right column) to Fromm's scheme.}
\end{figure}
An illustration of this is shown in Fig. \ref{fig:figerror} where the diffusion and dispersion errors are represented. These errors are quantified via a classical 
\textcolor{Reviewer2}{Von Neumann stability analysis (see \cite{Hirsch2007}, Sec.~7.4), also known as Fourier stability analysis. It consists in analysing the amplification factor of any harmonic of the signal, defined at any spatial position, as the ratio of the solutions at two successive time instants $G = u^k/u^{k-1}$. Through this procedure, }
the expression of the complex gain factor $G = \text{Re}(G) + \jmath \, \text{Im}(G)$ is extracted as function of the phase angle $\phi = m \Delta_x$ with $m$ corresponding to a spatial wavenumber. It is worth recalling that the phase angle $\phi=m \Delta_x = 2\pi / \lambda$, where $\lambda$ is the spatial wavelength, can also be written as $\phi = 2\pi / (N-1)$, where N represents the number of points used to discretize the signal over $\lambda$. 
To ensure stability, the diffusion error $\epsilon_D$, \textit{i.e.} the modulus of the gain factor $|G|$, should remain less than 1 for the whole range of $\phi$ present in the signal being advected. The quantity $1-|G|$ which represents the level of numerical diffusion must be minimized to avoid artificial decrease of wave amplitude components. The dispersive error is here characterized by $\epsilon_{\phi} =  \arctan(-\text{Im}(G)/\text{Re}(G)) / \sigma \phi $, which corresponds to the spurious multiplicative factor affecting the expected phase velocity of wave components. 
A good numerical scheme should keep the value of $\epsilon_{\phi}$ as close as possible to unity to limit phase advance or delay observed for $\epsilon_{\phi}>1$ or $\epsilon_{\phi}<1$, respectively. 
Figure \ref{fig:figerror} clearly shows the role played by $\delta$ on the diffusion and dispersion errors. 
The case $\delta=0$ (Lax Wendroff scheme) is characterized by a dominant phase lag within the stability bounds $0<\sigma<1$. The choice $\delta=\sigma (1-\sigma)/2$ (not shown) corresponds to  Beam and Warming scheme which yields a dominant phase advance error for $0<\sigma<1$. The case $\delta = \sigma (1-\sigma)/4$ corresponds to the famous Fromm's scheme which compensates to some extent the phase errors of the two aforementioned schemes for a wide range of $\sigma$. Indeed, we notice in Fig. \ref{fig:figerror} that isolines of $\epsilon_{\phi}$ lower than unity are significantly shifted towards the higher values of $\phi$, indicating that dispersive error levels can be \textit{a priori} significantly reduced in the intermediate range of $\phi$ corresponding to practical simulation cases. However, the reduction of diffusive errors, as illustrated with $|G|$ is far less efficient, in particular for high values of $\sigma$. The diffusive errors are even seen to increase for lower values of $\sigma$ and high values of $\phi$. 

We have to keep in mind that the MGEnKF algorithm employs ensemble members which have to be generated using relatively coarse grids (thus high values of $\phi$) for which both the dispersive and diffusive errors are likely to be important. The aforementioned observations of the non-monotonic and uncorrelated evolutions of $\epsilon_D$ and $\epsilon_{\phi}$ suggest that adjusting only $\delta$ is not sufficient to allow a satisfactory control of both kind of errors at the same time. 
An optimization of $\delta$ made for reducing the dispersive error could undesirably deteriorate the diffusive behavior of the scheme. 
This justifies that in the optimization strategy that we consider in Sec.~\ref{sec:advection}, we add to \eqref{eq:schema} an additional correction of the same order as the dispersive correction term in factor of $\delta$. 
This correction is chosen as being consistent with $\alpha \Delta_x^2 \dfrac{\sigma^2}{2}\dfrac{\partial ^2 u}{\partial x^2}$. With negative values of $\alpha$, this term will be expected to add an anti-diffusive behavior, counteracting the diffusion error intrinsically associated with the scheme \eqref{eq:schema}. The combined use of both correction terms is thus expected to allow a more relevant separated control of both dispersion and diffusion errors.

As previously observed in Fig. \ref{fig:figerror}, the properties of the numerical scheme significantly vary as function of $\phi$.
It is therefore far from being evident that considering $\delta$ and $\alpha$ as constant optimization parameters is sufficient to reduce the numerical error over a wide range of $\phi$ scales. 
In view of considering complex (spectrally richer) solutions and extending the use of this scheme to non-linear models, possibly leading to spatially evolving frequency content, it is thus also chosen to consider spatially varying functions for $\delta$ and $\alpha$ instead of constant values. This variability, which will be represented by expressing these parameters via spatial expansions of polynomials, will allow for local numerical optimization on coarse meshes.
 
\paragraph{Summary: numerical scheme retained to perform coarse-grid simulations}~\\
Following the \textit{a priori} analysis given all along this section, the following numerical scheme is finally retained:
\begin{equation}
\label{eq:1D_adv_oneparameter_parametrized}
\begin{aligned}
u_\text{j}^\text{k} = & \mathbf{\mathcal{M}}_{k:k-1}\left(u;\sigma, \delta \right) + \mathbf{\mathcal{C}}_{k:k-1}\left( u;\sigma, \alpha, \gamma \right) 
\end{aligned}
\end{equation}
\textcolor{Reviewer2}{
where
\begin{equation}
\begin{aligned}
\mathbf{\mathcal{M}}_{k:k-1}\left(u;\sigma, \delta \right)& = & u_\text{j}^\text{k-1}-\frac{\sigma}{2} \left( u_\text{j+1}^\text{k-1}-u_\text{j-1}^\text{k-1} \right) + \frac{\sigma^2}{2} \left( u_\text{j+1}^\text{k-1}-2 u_\text{j}^\text{k-1}+u_\text{j-1}^\text{k-1} \right) \\
& & +\delta \left( -u_\text{j-2}^\text{k-1}+3u_\text{j-1}^\text{k-1}-3u_\text{j}^\text{k-1}+u_\text{j+1}^\text{k-1} \right), 
\end{aligned}
\label{eq:1D_adv_oneparameter_parametrized_M}
\end{equation}
and 
\begin{equation}
\mathbf{\mathcal{C}}_{k:k-1}\left( u;\sigma, \alpha, \gamma \right)  =
\alpha\frac{\sigma^2}{2} \left( u_\text{j+1}^\text{k-1}-2 u_\text{j}^\text{k-1}+u_\text{j-1}^\text{k-1} \right)+\gamma \left( -u_\text{j-2}^\text{k-1}+3u_\text{j-1}^\text{k-1}-3u_\text{j}^\text{k-1}+u_\text{j+1}^\text{k-1} \right) 
\label{eq:1D_adv_oneparameter_parametrized_C}
\end{equation}
}
Here, one can see that the optimization of $\delta$ is not performed directly, but via a parameter $\gamma$ which measures the deviation of the optimized dispersion coefficient from the constant value $\delta=\sigma(1-\sigma^2)/4$ proposed by Fromm. This choice has been performed to provide a clear separation between the dynamical model $\mathbf{\mathcal{M}}$ and the correction model $\mathbf{\mathcal{C}}$ when comparing \eqref{eq:schema} and \eqref{eq:1D_adv_oneparameter_parametrized}. The variability in space of the coefficients $\gamma$ and $\alpha$ is obtained expressing them in terms of Legendre Polynomial expansions truncated to the order $n$:
 \begin{equation}
 \gamma\left(x\right)=\gamma_\text{0}P_\text{0}\left(x\right)+\gamma_\text{1}P_\text{1}\left(x\right)+\cdots+\gamma_\text{n}P_\text{n}\left(x\right)\label{eq:beta_legendre},
\end{equation}
and
\begin{equation}
\alpha\left(x\right)=\alpha_\text{0}P_\text{0}\left(x\right)+\alpha_\text{1}P_\text{1}\left(x\right)+\cdots+\alpha_\text{n}P_\text{n}\left(x\right)\label{eq:alpha_legendre}.
\end{equation}

Preliminary tests showed that $4$-th order representation ($n=4$) is satisfactory for the cases considered in this study. The inner loop will be used to optimize the expansion coefficients for a total of $10$ parameters (five expansion coefficients $\gamma_i$ and five expansion coefficients $\alpha_i$). The values for these parameters could possibly be constrained during the inner loop optimization in order to accept only values leading to stable solutions. In particular, the extended stability constraint for the present scheme in absence of additional anti-diffusive correction is imposed, which reads as
:
\begin{equation}
\label{stab}
\gamma (1-2\sigma) + \frac{1}{4} \sigma^2 (1-\sigma^2) \geq 0
\end{equation}


\section{Application: one-dimensional advection equation}
\label{sec:advection}

The sensitivity of the MGEnKF to the performance of the \emph{inner loop} introduced in Sec.~\ref{sec:maths} \textcolor{Reviewer2}{is now assessed in a }practical DA experiment. More precisely, the \emph{inner loop} is here used to optimize the behavior of the numerical models for the one-dimensional linear equation discussed in Sec.~\ref{sec:apriorianalysis}. The optimization, which is performed over the polynomial expansion coefficients $\gamma_i$ and $\alpha_i$, aims to reduce the numerical error of the ensemble simulations which are performed on coarse grids.   

\subsection{Set-up of test case and test solutions on coarse meshes\label{sec:adv_coarse_model}}
\begin{figure}[htp]
\centering
\includegraphics[width=1\textwidth]{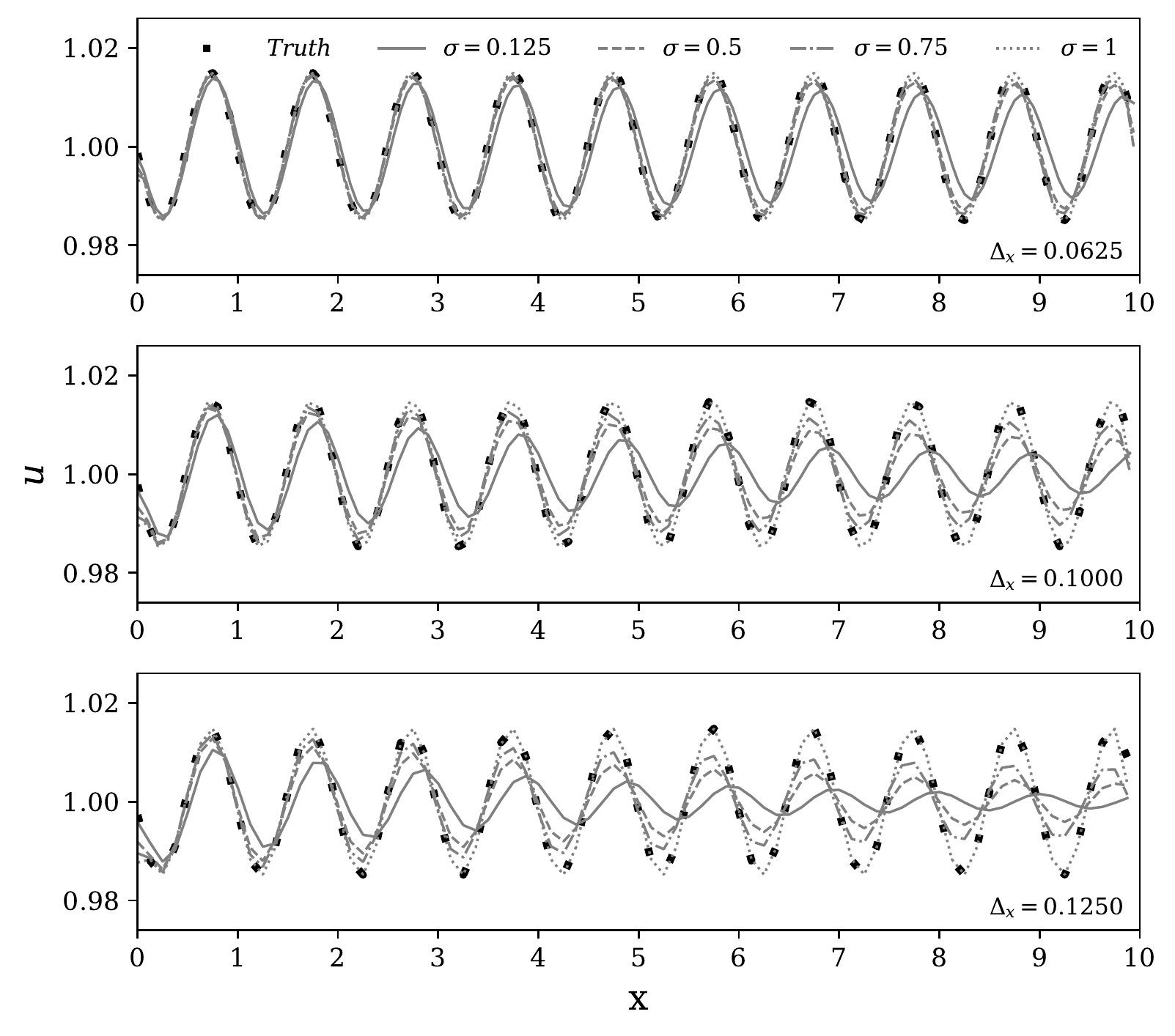}
\caption{\label{fig:1D-advection_coarse}Preliminary simulations for the 1D advection equation. Solutions at $t=10$ for different grid refinement and $\sigma$ values are compared with the \emph{true} state.}
\end{figure}

The set-up of the test case representing the one-dimensional linear advection equation (see \eqref{eq:1D_adv} in Sec. \ref{sec:apriorianalysis}) is now presented. The constant advection velocity is set to $c=1$.
Preliminary numerical tests are carried out with the scheme presented in \eqref{eq:1D_adv_oneparameter_parametrized} and by setting \textit{a priori} $\alpha=0$ and $\gamma=0$ (Fromm scheme). The initial condition is set to $u(x,t=0)=c$ everywhere in the physical domain. 
A Dirichlet time-varying condition is imposed at the inlet:
\begin{equation}
u(x=0,t)=c \, \left(1 +\theta \sin\left(2\pi t\right) \right),
\label{eq:inlet-1D-advection}
\end{equation}
where $\theta$ represents the amplitude of a sinusoidal perturbation of period $T = 1$ and is set to $\theta=0.015$. \textcolor{Reviewer2}{Every time-step, the outlet boundary condition is extrapolated along grid-lines from the interior nodes to the boundary points at $x=10$ using 4-th order Lagrange polynomials. This represents a classical outlet condition for simulating advective flows (see Sec. 8.10.2 in \cite{Ferziger2002_springer}), that can be used to direct the flow outside of the physical domain}. The simulations are performed over a computational domain of size $0 \leq x  \leq 10$ in $L_0= c T$ units. Three different levels of mesh refinement (moderate to low) are chosen for these tests. The resolution is chosen to be of practical interest for usage for the ensemble members in the MGEnKF algorithm. The mesh size is set with constant values of $\Delta_x=0.0625$, $0.1$ and $0.125$, respectively. This corresponds to $16$, $10$ and $8$ discrete nodes per characteristic length $L_0$ or equivalently, phase angles around $\phi=0.4$, $0.6$ and $0.8$. According to Fig. \ref{fig:figerror}, relatively moderate error levels can be observed with these resolution levels. However, their accumulation during the signal advection is expected to become significant.
The preliminary simulations are also performed using different CFL numbers $\sigma$. 
The results, which are shown in  Fig. \ref{fig:1D-advection_coarse}, are compared with the \emph{true} (exact) known solution. One can see that, with the exception of the finer mesh resolution and higher values of $\sigma$, the numerical solutions are rapidly affected by the accumulation of diffusive and dispersive errors. In particular, a significant amplitude reduction and phase advance can be observed. It is worth recalling that these errors could be naturally eliminated in the present case for the specific choice of $\sigma=1$. However, this constraint is not necessarily compatible with practical needs associated with the numerical simulations running within the MGEnKF algorithm (reduced $\sigma$ required to ensure stability with more complex boundary conditions, signal synchronization with observation, and so on).

\subsection{Performance of the MGEnKF algorithm without the \emph{inner loop}\label{sec:adv_coarse_MGENKF}}
The performance of the MGEnKF algorithm without the \emph{inner loop} may be severely degraded by the numerical errors induced by the use of coarse grids for the ensemble members. This can be shown with a simple twin-experiment where observation is accessible relatively far from the inlet and the DA tool attempts to estimate the inlet parameter $\theta$. 
Data assimilation is performed with the following conditions:
\begin{itemize}
\item Observations are generated from the analytical solution with $\theta=0.015$ on the space domain $[3, 4]$ and on the time window $[0, 390]$. The sampling frequency is set so that approximately $15$ observation updates per characteristic evolution time $L_0 / c$ are obtained, for a total of $\approx 6000$ DA analysis phases. Also, the time origin for the sampling is shifted of ten characteristic times so that the state for $t=0$ is \emph{fully developed}, \textit{i.e.} the initial condition $u(x,0)=c$ is completely advected outside the computational domain. For simplicity, we assume that the observations and the coarse-grid ensemble are represented on the same space. Therefore, $\mathbf{\mathcal{H}}_\text{k}\equiv\mathbf{\mathcal{H}}$ is a subsampling operator independent of time retaining only the points comprised in the coarse space domain $[3,4]$.
The observations are artificially perturbed using a constant in time Gaussian noise of diagonal covariance $\mathbf{R}_\text{k}\equiv\mathbf{R}=2.25\cdot 10^{-6}\mathbf{I}$. This choice has been performed for every test case following the recommendations of \cite{Tandeo_Ailliot_Bocquet_Carrassi_Miyoshi_Pulido_Zhen_MWR_2020}, which extensively investigated the sensitivity of the EnKF to the noise / uncertainty in the model and in the observation.

\item The \emph{model} is represented by \textit{i)} a main simulation with a resolution of $\Delta_x=0.0125$ (\textit{i.e.} $80$ mesh elements per $L_0$) and \textit{ii)} ensemble simulations performed on coarse grids. Multiple runs of the MGEnKF are performed using three different mesh resolutions for $\Delta_x$ (equal to $0.125$, $0.1$ and $0.0625$) for the ensemble members and also imposing different values for $\sigma$. The \emph{model} employs fixed parameters $\alpha=0$ and $\gamma=0$ (Fromm scheme) for all cases. The size of the ensemble is set to $N_\text{e}=100$. The amplitude of the sinusoidal inlet perturbation $\theta$ for the ensemble simulations is considered to be unknown. It is initially assumed to be described by Gaussian distribution $\theta \sim \mathcal{N}(0.025, \mathbf{Q}_{\theta})$, with $\mathbf{Q}_{\theta}(t=0)=2.5 \cdot 10^{-7}\mathbf{I}$. For the main simulation run on the fine grid, the mean value of the Gaussian distribution, \textit{i.e.} $\theta=0.025$ is initially imposed. These values are significantly far from $\theta=0.015$ used with the analytical solution. This choice allows to analyze the rate of convergence of the optimization procedure. The initial condition $u(x,t=0)=c$ is used for the main simulation on the fine-grid as well as for the coarse ensemble simulations. It is worth recalling that, in such a case, at $t=0$, the true state exhibits a very different solution. This choice allows to check the robustness of the algorithm during the transient solution and ascertain the correct evolution of the first state estimation stages when the solution of the model may be very different from the observations. 
\end{itemize}

\begin{figure}[htbp]
\includegraphics[width=1\textwidth]{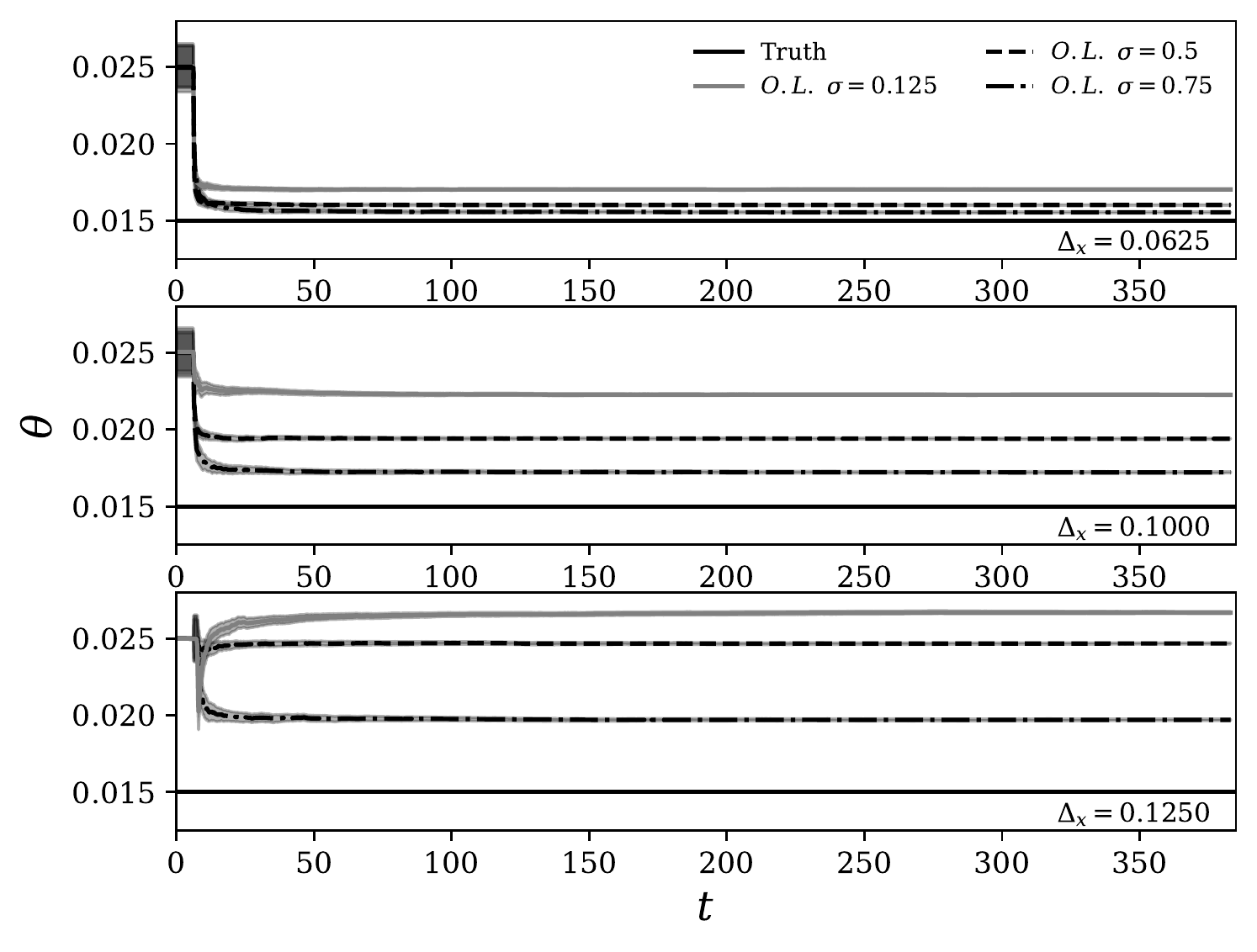}
\caption{\label{fig:amplitude_adv_OL}
Estimation history of the inlet parameter $\theta$ using the MGEnKF without the \emph{inner loop}. The DA method is performed using three different mesh resolutions for the ensemble members and varying the parameter $\sigma$.}
\end{figure}

\begin{figure}[htbp]
\centering
\includegraphics[width=1\textwidth]{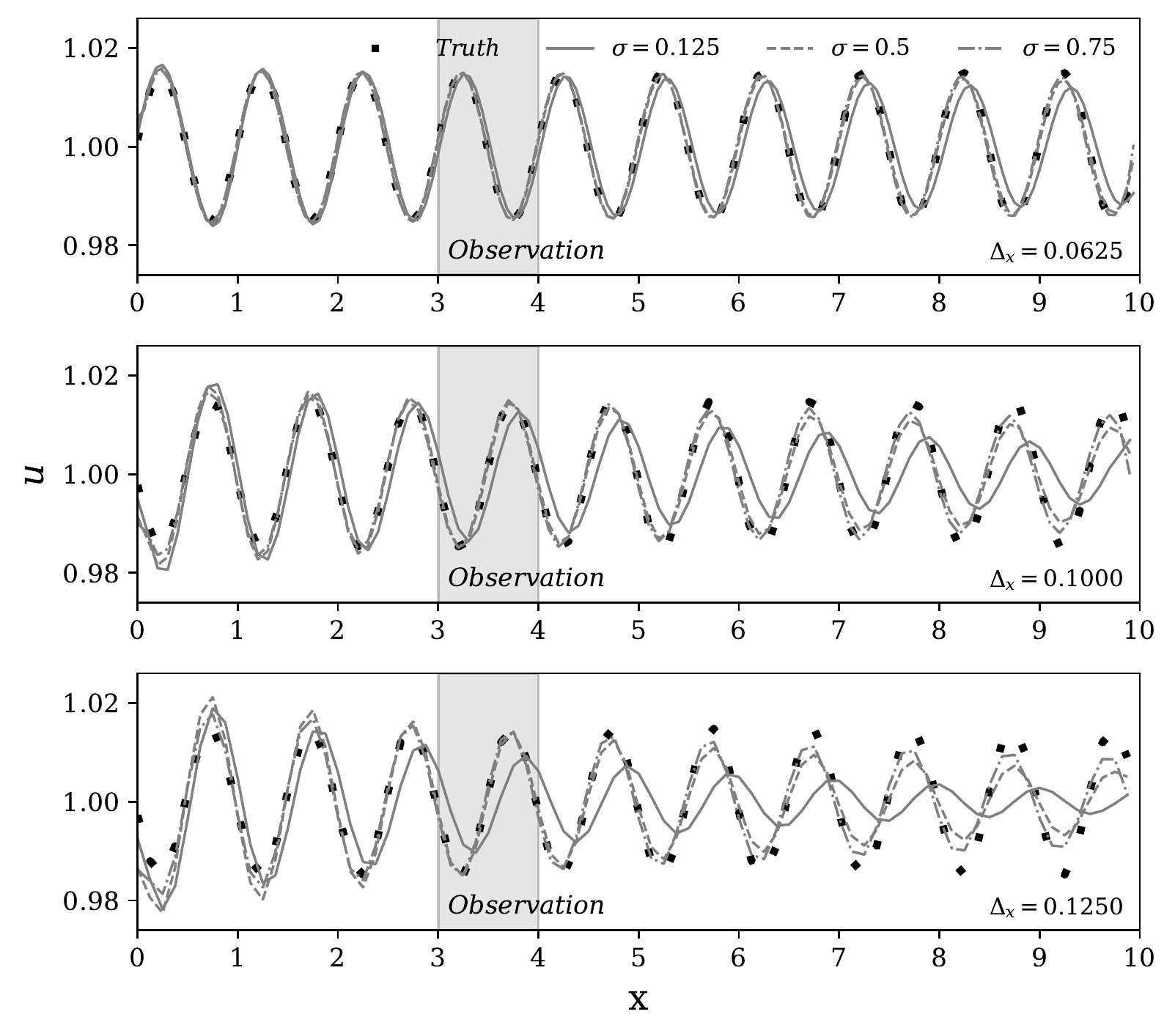}
\caption{\label{fig:1D-advection_MGENKF}State estimation obtained for the ensemble members via MGEnKF without the \emph{inner loop} for the linear advection equation test case. Comparisons with the exact solution are shown for $t=300$ for different grid refinement levels and $\sigma$ values.}
\end{figure}

The time-evolution of the estimation of $\theta$ is shown in Fig. \ref{fig:amplitude_adv_OL}. First of all, one can see that the estimation procedure starts at $t=10$. This choice is consistent with the positioning of the sensors for observation, which are located in the middle of the computational domain ($[3, 4]$).
The estimation of $\theta$ is progressively degraded as the refinement is decreased, with the most accurate results obtained with the finest mesh refinement $\Delta_x=0.0625$ for any given $\sigma$. Concerning the influence of the CFL number $\sigma$, the model prediction is generally more accurate as $\sigma \to 1$. One can also see that progressively larger deviations are observed varying $\sigma$ for coarser meshes.
Therefore, the increased dispersive and diffusive errors associated with lower CFL numbers are the cause for the large deviations observed when $\sigma=0.125$. $\theta$ is naturally over-estimated in this case as the calculations of the model in the sampling region are dominated by numerical errors. The amplitude of the sinusoidal wave imposed at the inlet is numerically \emph{diffused} and \emph{dispersed} between $0\leq x \leq 3$. However, the optimized value of $\theta$ determined by MGEnKF compensates for the mismatch between model and reference in the sampling region. This can be clearly observed in Fig. \ref{fig:1D-advection_MGENKF}, where the ensemble coarse-grid estimation is shown. 

\subsection{Performance of the MGEnKF with the \emph{inner loop}}\label{sec:adv_coarse_oMGENKF}

In this section, the complete MGEnKF scheme is used to study the same DA problem investigated in Sec. \ref{sec:adv_coarse_MGENKF}. This analysis will allow to unambiguously identify the contribution of the \emph{inner loop} for the optimization of the ensemble members running on the coarse grid level. 


The main modification when compared with the previous analysis is that now the parameters $\gamma(x)$ and $\alpha(x)$ in \eqref{eq:1D_adv_oneparameter_parametrized} are considered to be unknown space varying model parameters which will be optimized using the \emph{inner loop}. The additional term related to the correction model $\mathcal{C}_{k:k-1}$ can be explicitly written as $\alpha\frac{\sigma^2}{2} \left( u_\text{j+1}^\text{k-1}-2 u_\text{j}^\text{k-1}+u_\text{j-1}^\text{k-1} \right) + \gamma \left( -u_\text{j-2}^\text{k-1}+3u_\text{j-1}^\text{k-1}-3u_\text{j}^\text{k-1}+u_\text{j+1}^\text{k-1} \right)$, see \eqref{eq:1D_adv_oneparameter_parametrized_C}. In particular, the optimization will target the values of the Legendre polynomial expansion coefficients $\gamma_i$ and $\alpha_i$ introduced in \eqref{eq:beta_legendre} and \eqref{eq:alpha_legendre} using the fine-grid state as surrogate observation.

The MGEnKF thus performs two optimization procedures within the analysis phase, one in the \emph{inner loop} and a second one in the \emph{outer loop}:
\begin{enumerate}
\item Optimization of the polynomial expansion coefficients $\alpha_i$ and $\gamma_i$ to reduce the discrepancy between the low-fidelity (ensemble members) and high-fidelity (main simulation) models in the \emph{inner loop}.
\item Optimization of the amplitude of the inlet perturbation $\theta$ in the \emph{outer loop}.
\end{enumerate}
The coefficients $\alpha_i$ and $\gamma_i$ are initially described by Gaussian distributions $\alpha_\text{i} \sim \mathcal{N}(0, \mathbf{Q})$, $\gamma_\text{i} \sim \mathcal{N}(0, \mathbf{Q})$ with $\mathbf{Q}(t=0)=9\cdot 10^{-8}\mathbf{I}$ and $i=0, 1, 2, 3, 4$. The information from the entire fine-grid domain is available. Therefore,  $\mathbf{\mathcal{H}}_\text{k}^{\text{o}}\equiv\mathbf{\mathcal{H}}^{\text{o}}$ is a subsampling operator independent of time retaining all the points comprised in the coarse space domain $[0,10]$. Preliminary tests showed that, in order to improve the performance of the \emph{inner loop}, the surrogate observation from the fine grid should be perturbed using a constant in time Gaussian noise of covariance $\mathbf{R}^\text{o}_\text{k}\equiv\mathbf{R}^\text{o}=1\cdot 10^{-8}\mathbf{I}$. The value here used for $\mathbf{R}^\text{o}$ is orders of magnitude lower than the observation covariance matrix $\mathbf{R}$. Therefore, one can consider the surrogate observation as a quasi-exact observation.
It should be noted that the EnKF procedure in the \emph{inner loop} only optimizes the parameters affecting the model term $\mathcal{C}$ included in the low-fidelity model and no state estimation is performed on the coarse ensemble within this phase.

\begin{figure}[htbp]
\includegraphics[width=1\textwidth]{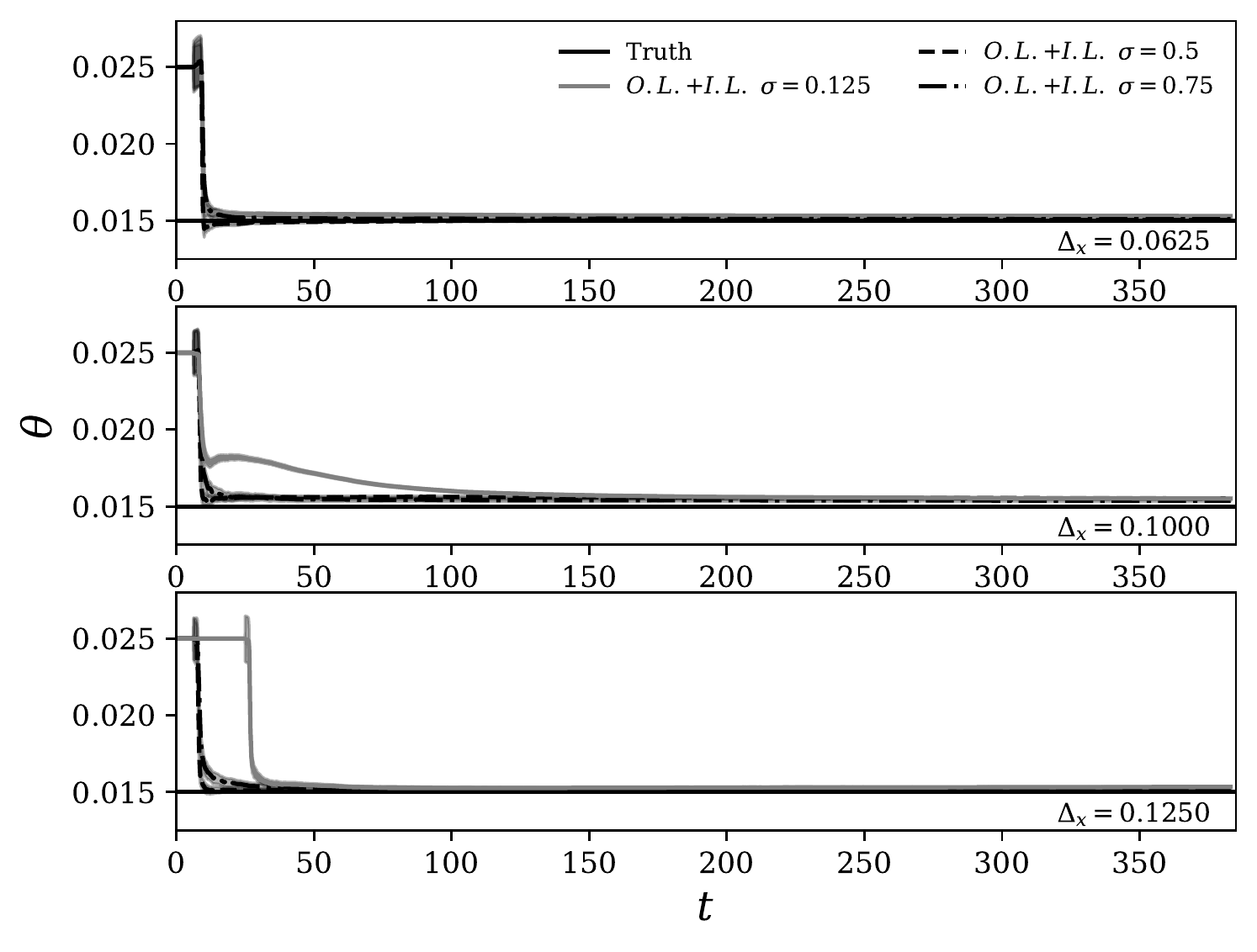}
\caption{\label{fig:amplitude_adv_optimized}
Estimation history of the inlet parameter $\theta$ using the complete MGEnKF. The DA method is performed using three different mesh resolutions for the ensemble members and varying the parameter $\sigma$.}
\end{figure}

The estimation of the parameter $\theta$ using the complete MEnKF is shown in Fig. \ref{fig:amplitude_adv_optimized}. As in Sec. \ref{sec:adv_coarse_model}, runs have been performed using three levels of mesh refinement for the ensemble members. The first ten characteristic times of the experiment are used to initialize the tuning of the parameters $\alpha$ and $\gamma$ (\textit{i.e.} \emph{outer loop} initially deactivated). For $t>10$, both optimization procedures are performed. During the very first phases of the assimilation process ($10<t<20$), the estimated value of $\theta$ gets within the $5\%$ error margin when compared to the \emph{truth} and no degradation is observed for every mesh refinement / $\sigma$ combination investigated. The convergence is noticeably slower when $\sigma=0.125$ for $\Delta_x=0.100$ and $\Delta_x=0.125$, where the numerical errors in the initial phase are the largest. For the worst case scenario ($\sigma=0.125$, $\Delta_x=0.125$), an initial phase of $40$ characteristic times was required to obtain converged results for the inner loop, which delayed the start of the outer loop.
Overall, the complete MGEnKF outperforms the version without the \emph{inner loop}. This result highlights the complementary features of the two optimization strategies to obtain a global accurate representation of the flow.

\begin{figure}[htbp]
\includegraphics[width=1\textwidth]{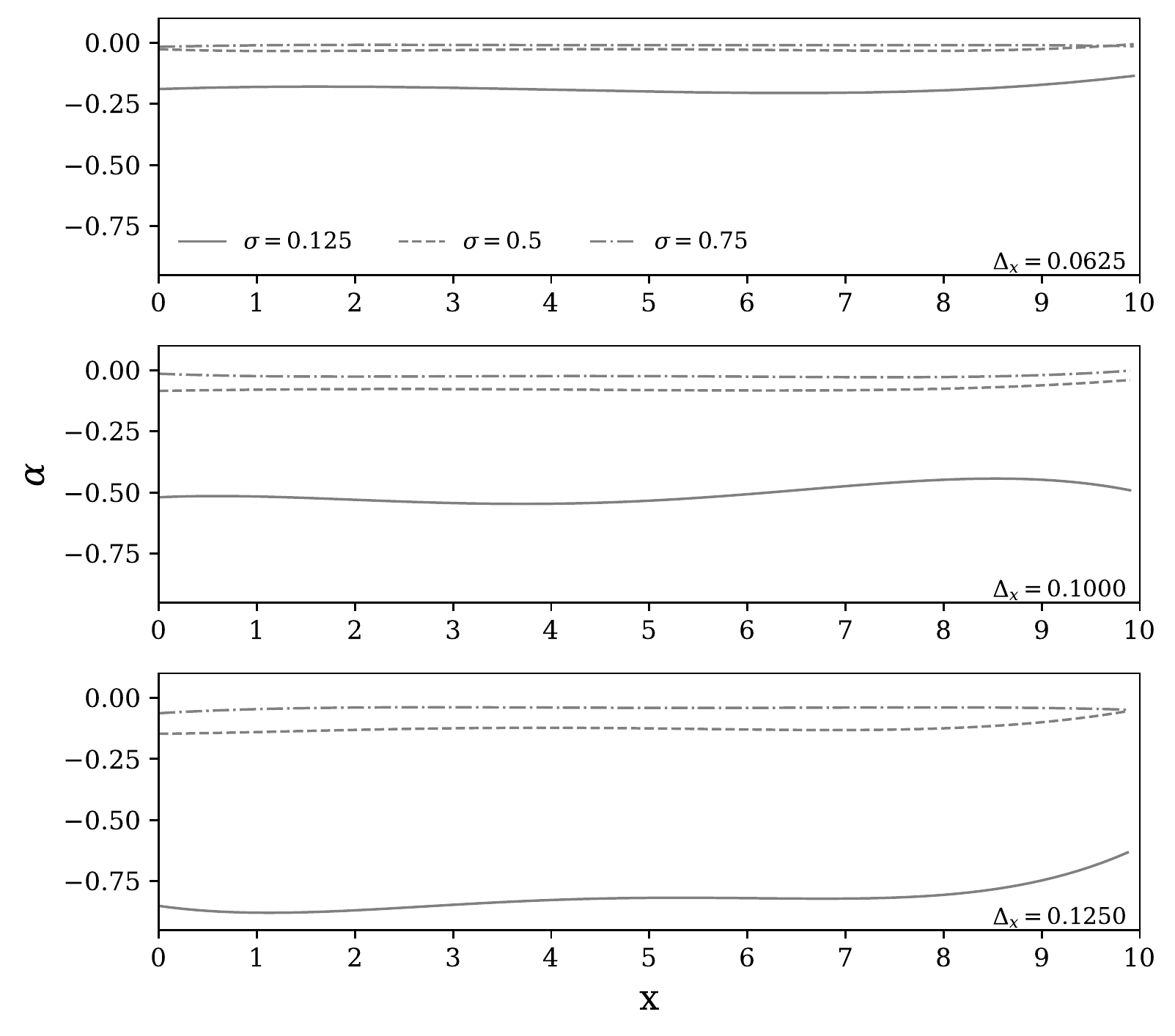}
\caption{\label{fig:adv_alpha}
Values of the parameter $\alpha$ obtained via the \emph{inner loop}. Results are shown for different grids and $\sigma$ values for a simulation time $t=300$.}
\end{figure}

The optimization performed in the inner loop is now analyzed in detail. The parameter $\alpha(x)$ is shown in Fig. \ref{fig:adv_alpha} at $t=300$. 
As expected, one can see that $\alpha$ exhibits negative values which approach zero with increasing mesh resolution and higher $\sigma$ values. The numerical diffusion observed in the model when computed on coarser meshes increases, thus the optimization procedure provides an anti-diffusive contribution.

\begin{figure}[htbp]
\includegraphics[width=1\textwidth]{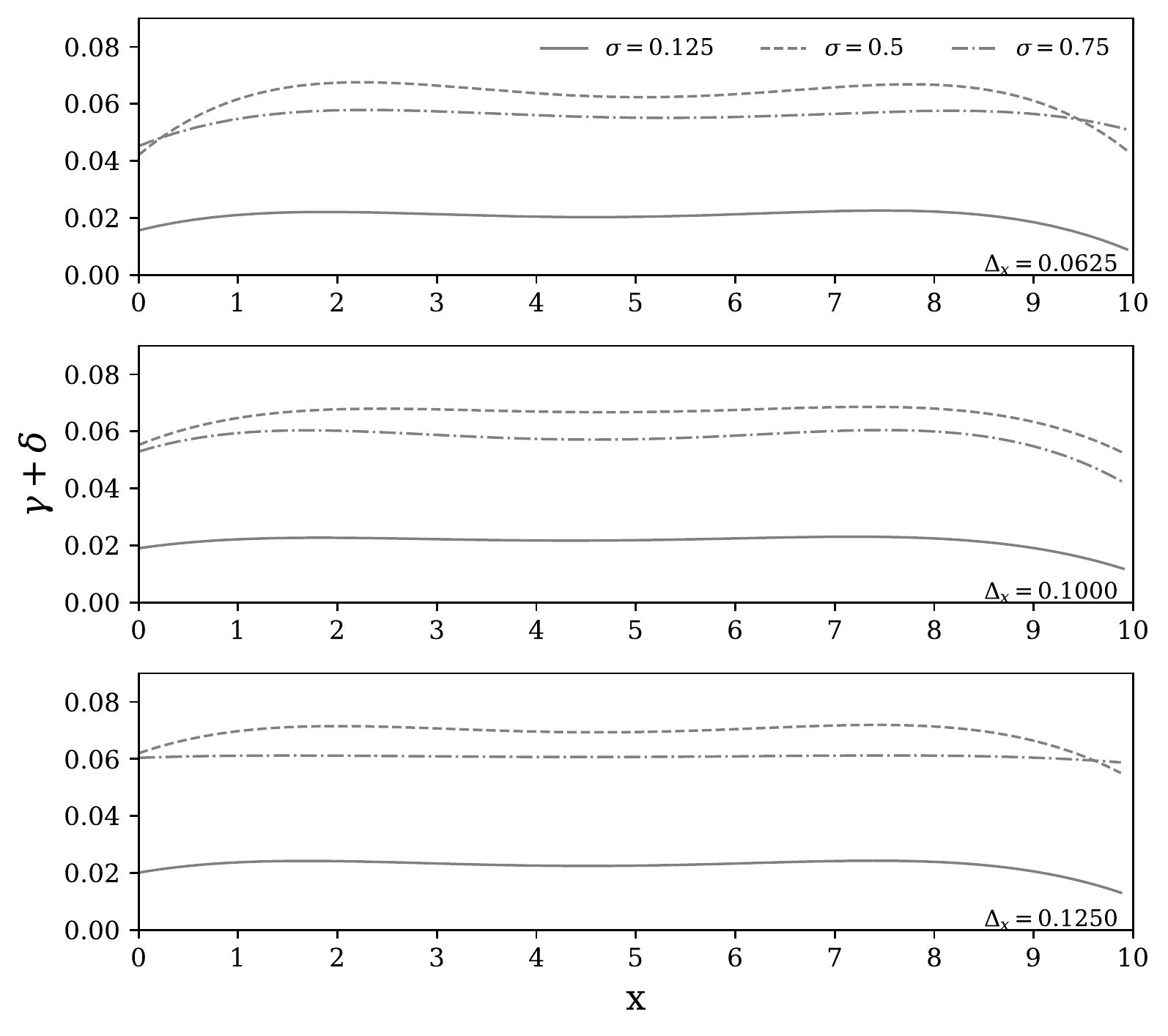}
\caption{\label{fig:adv_gamma}
Values of the parameter $\gamma + \delta$ obtained via the \emph{inner loop}. Results are shown for different grids and $\sigma$ values for a simulation time $t=300$.}
\end{figure}

The spatial distribution of the sum of the parameters $\gamma + \delta$ representing the total dispersion of the scheme \eqref{eq:1D_adv_oneparameter_parametrized} is presented in Fig. \ref{fig:adv_gamma} at $t=300$. For every case analyzed, one can remark that $\gamma + \delta$ tend to converge towards the value for which the scheme \eqref{eq:1D_adv_oneparameter_parametrized} becomes third order accurate, that is $\delta + \gamma=\sigma\left(1-\sigma^2\right)/6$ ($0.0547$, $0.0625$ and $0.0205$ for $\sigma$ equal to $0.75$, $0.5$ and $0.125$, respectively). This result is expected since this particular value cancels out the dominant dispersive error in the scheme.

\begin{figure}[htbp]
\centering
\includegraphics[width=1\textwidth]{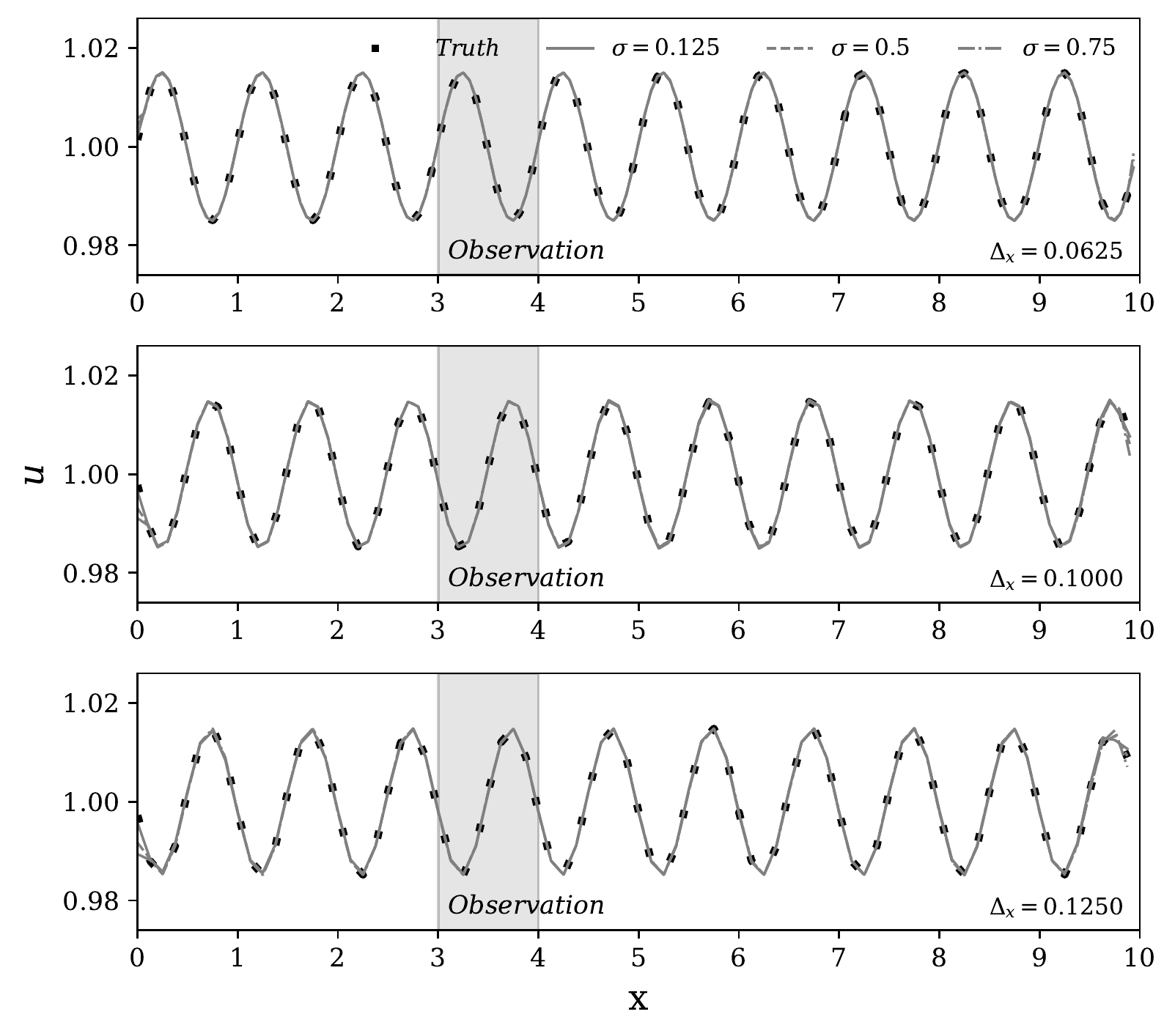}
\caption{\label{fig:1D-advection_OMGENKF}Solutions provided by the ensemble members in the complete MGEnKF. Results, which are obtained for different meshes and values of $\sigma$, are compared with the \emph{true} state at $t=300$.}
\end{figure}

Finally, results for the ensemble members are shown in Fig. \ref{fig:1D-advection_OMGENKF} at $t=300$. One can see a marked improvement when these results are compared to the ones shown in Fig. \ref{fig:1D-advection_MGENKF}. The estimation of the parameter $\theta$ is clearly much more accurate ($5\%$ error) and there is virtually no difference in prediction between the \emph{truth} and model used on the coarse grids. This result has been obtained owing to the suppression of the numerical diffusion and dispersion errors via \emph{inner loop}, which proved to be efficient for every configuration analyzed (wide range of phase angle / CFL numbers). DA analyses considering more complex inlet conditions (multiple frequencies $\theta_i$) have been performed to assess the method. The results, which are not presented here for sake of brevity, show similar accuracy.

\section{Application: one-dimensional viscous Burgers' equation}
\label{sec:Burgers}

Let us now consider the non-linear and viscous 1D Burgers' equation:
\begin{equation}
\frac{\partial u}{\partial t}+u\frac{\partial u}{\partial x}=\nu\frac{\partial^2 u}{\partial x^2}\label{eq:1D_burgers}
\end{equation}
where $x$ is the spatial coordinate, $u$ the velocity and $\nu$ the kinematic viscosity. Considering a centered difference scheme for both the convection and the diffusion term over a uniform grid with mesh size $\Delta_x$, and an explicit forward first order scheme for the time derivative, one obtains:

\begin{align}
u^{\text{k}}_{\text{j}}=&u^{\text{k-1}}_{\text{j}}-u^{\text{k-1}}_{\text{j}}\frac{\Delta_t}{2\Delta_x}\left(u^{\text{k-1}}_{\text{j+1}}-u^{\text{k-1}}_{\text{j-1}}\right)+\nu\frac{\Delta_t}{{\Delta_x}^2}\left(u^{\text{k-1}}_{\text{j+1}}-2u^{\text{k-1}}_{\text{j}}+u^{\text{k-1}}_{\text{j-1}}\right) \label{eq:1D_burgers_scheme}
\end{align}
where $\Delta_t$ represents the time step. 

Similarly to what was done in Sec.~\ref{sec:advection}, the performance of the MGEnKF is studied. However, owing to the non-linearity of \eqref{eq:1D_burgers}, a model of the numerical error associated with the discretization process cannot be derived from the dynamical equation. Thus, for this case, the optimization by the inner loop is performed using the \textcolor{Reviewer2}{same dispersion correction term as in \eqref{eq:schema}}. While this model has not been derived for the dynamical equations \eqref{eq:1D_burgers}-\eqref{eq:1D_burgers_scheme}, one can assess the degree of precision attained in reducing amplitude and phase errors.

A numerical experiment for this test case is first performed using a high-resolution mesh to obtain a reference solution and to generate observation for the MGEnKF application. A Dirichlet time-varying condition is imposed at the inlet:
\begin{equation}
u(x=0,t)=u_0 \left(1+\theta\sin\left(2\pi t\right) \right),
\label{eq:inlet-1D-burgers}
\end{equation}
where $u_0=1$ is the mean characteristic velocity of the flow and $\theta$ represents the amplitude of a sinusoidal signal whose period is $T =1$. The amplitude parameter has been set to $\theta=0.2$ in order to observe significant non-linear effects with the Reynolds number chosen for this application, which will be discussed in the following. \textcolor{Reviewer2}{As described in the previous section, the outlet boundary condition is extrapolated along grid-lines from the interior nodes to the boundary points at $x=10$ using 4-th order Lagrange polynomials at every time step.} The initial condition is $u(x,t=0)=u_0$ everywhere in the physical domain. The Reynolds number is set to $Re=\frac{u_0 \, L_0}{\nu}=200$, where $L_0 =u_0 \, T$ is the mean wave-length of the signal and the characteristic length of the system. All physical lengths characterizing the system are normalized by $L_0$. The simulation is performed over a computational domain of size $10$ length units and the origin of the system is set so that $0 \leq x  \leq 10$. 
The mesh resolution is set to $64$ discrete nodes per $L_0$, for a total of $640$ mesh elements. The constant time step $\Delta_t$ is set so that the mean CFL number is $CFL=\frac{u_0 \Delta_t}{\Delta_x}=0.025$, which is small enough to guarantee a stable numerical evolution of the system. 

\begin{figure}[htbp]
\centering
\includegraphics[width=1\textwidth]{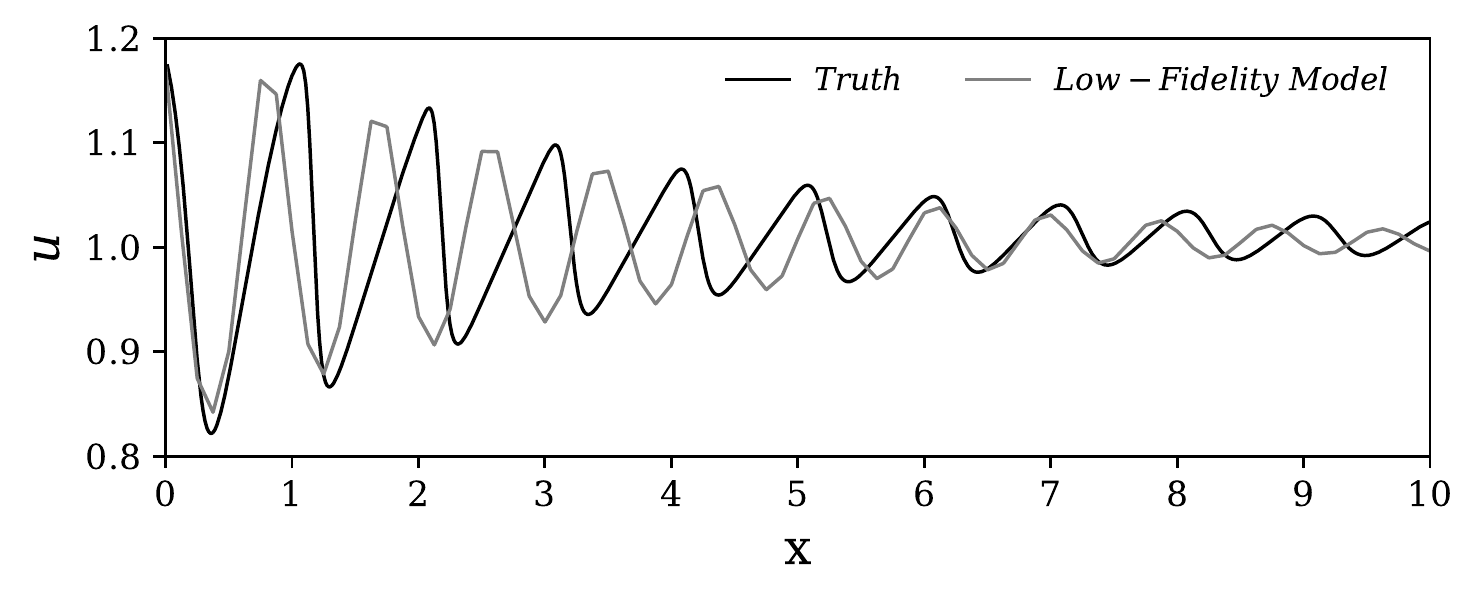}
\caption{\label{fig:1D-burgers_CGmodel}Instantaneous solution of the 1D Burgers' equation at $t=100$. Solutions obtained via a very refined simulation (Truth, black line) and using a coarse grid (Low-Fidelity Model, gray line) are compared.}
\end{figure}

The predicted solution using this model is referred to as the \emph{true} state of the system. This state is first compared with the prediction obtained via a  low-fidelity model, which is identical to the reference simulation but uses only $8$ nodes per length $L_0$, for a total of $80$ mesh elements. The comparison of the two solutions, which is shown in Fig. \ref{fig:1D-burgers_CGmodel}, clearly indicates that the lack of mesh resolution is responsible for important errors in the amplitude and in the phase of the velocity signal. 
In particular, the main source of numerical error appears to be of dispersive nature. More specifically, the time period of a full oscillation of the velocity field is significantly shorter when compared with the reference simulation. Diffusion errors are also visible, although their magnitude is smaller. The combination of these two sources of error severely affects the representation of the non linear phenomena at play. In fact, in the reference simulation, one can see that non-linear dynamics are strong enough to sensibly deform the sinusoidal profiles imposed at the inlet. On the other hand, the state predicted via the low-fidelity model does not show marked deformations of the velocity profile, suggesting that non-linear effects are poorly represented.

In order to perform an extensive test of the performance of the MGEnKF strategy, two different runs are performed. The first one includes an outer loop only, while the second one performs the complete inner loop and outer loop scheme. This comparison will allow to assess the impact of the inner loop performance over the optimization of the global coefficients of the simulation.
Similarly to what is proposed in Sec.~\ref{sec:advection}, the numerical scheme given by \eqref{eq:1D_burgers_scheme} is modified to introduce \emph{model correction} terms:

\begin{align}
u^{\text{k}}_{\text{j}}=&u^{\text{k-1}}_{\text{j}}-u^{\text{k-1}}_{\text{j}}\frac{\Delta_t}{2\Delta_x}\left(u^{\text{k-1}}_{\text{j+1}}-u^{\text{k-1}}_{\text{j-1}}\right)+\nu\frac{\Delta_t}{{\Delta_x}^2}\left(u^{\text{k-1}}_{\text{j+1}}-2u^{\text{k-1}}_{\text{j}}+u^{\text{k-1}}_{\text{j-1}}\right)+\mathbf{\mathcal{C}}_{k:k-1}(u;\alpha,\gamma) \nonumber \\=&u^{\text{k-1}}_{\text{j}}-u^{\text{k-1}}_{\text{j}}\frac{\Delta_t}{2\Delta_x}\left(u^{\text{k-1}}_{\text{j+1}}-u^{\text{k-1}}_{\text{j-1}}\right)+\left(1+\alpha\right)\nu\frac{\Delta_t}{{\Delta_x}^2}\left(u^{\text{k-1}}_{\text{j+1}}-2u^{\text{k-1}}_{\text{j}}+u^{\text{k-1}}_{\text{j-1}}\right) \nonumber \\  &+\gamma \left(-u^{\text{k-1}}_{\text{j-2}}+3u^{\text{k-1}}_{\text{j-1}}-3u^{\text{k-1}}_{\text{j}}+u^{\text{k-1}}_{\text{j+1}}\right). \label{eq:1D_burgers_scheme_modified}
\end{align}
The model $\mathbf{\mathcal{C}}$ here introduced is composed by two correction terms which are driven by the parameters $\alpha(x,t)$ and $\gamma(x,t)$. $\alpha$ and $\gamma$ are identically zero in the main simulation of the MGEnKF, while they are optimized in the inner loop for the ensemble members. First, the $\alpha$ parameter controls a diffusive effect / numerical viscosity term. For this reason, local values are bounded to respect the condition $\alpha(x,t) \geq -1$,  \textit{i.e.} non physical solutions with negative global viscosity are excluded. Figure \ref{fig:1D-burgers_CGmodel} shows that grid coarsening is responsible for an over estimation of diffusive effects. Therefore, one should expect to observe a convergence of the parameter $\alpha(x,t)$ towards negative values. The second correction term $\gamma \left(-u^{\text{k-1}}_{\text{j-2}}+3u^{\text{k-1}}_{\text{j-1}}-3u^{\text{k-1}}_{\text{j}}+u^{\text{k-1}}_{\text{j+1}}\right)$ mimics the effects of a dispersion term of the form $\left(\gamma{\Delta_x}^3u_{xxx}\right)$ \cite{Hirsch2007}. This term was used in the previous section to correct the dispersive errors observed in the advection equation. The time evolution of $\alpha(x,t)$ and $\gamma(x,t)$ is taken into account by the MGEnKF itself, as the parameters are updated at each inner analysis phase. On the other hand, the space variability of the two parameters is obtained expressing them in terms of a Legendre Polynomial expansion. 
Similarly to what was done for the linear advection case presented in Sec.~\ref{sec:advection}, the expansion is truncated to $n=4$, \textit{i.e.} a $4$-th polynomial order. This implies that the optimization performed in the \emph{inner loop} targets the value for the ten model coefficients $\gamma_\text{i}$ and $\alpha_\text{i}$.

The performance of the estimators (inner and outer loop, outer loop only) is assessed via the following data-assimilation strategy:
\begin{itemize}
\item The observations are sampled each $160$ time steps of the reference simulation run on the space domain $[3, 4]$ ($64$ sensors) and on the time window $[0, 240]$. Considering the value of the time step $\Delta_t$ employed for the investigation, this implies that approximately $20$ analysis phases per characteristic time evolution $T$ are performed. The sampling of the reference simulation is performed starting from a \emph{fully developed} state at $t=0$. This is easily done owing to the periodic characteristics of the inlet. For sake of simplicity, we assume that the observations and the coarse-grid ensemble are represented on the same space. Therefore, $\mathbf{\mathcal{H}}_\text{k}\equiv\mathbf{\mathcal{H}}$ is a sub-sampling operator independent of time retaining only the points comprised in the coarse space domain $[3,4]$.
The observations are artificially perturbed adding a constant in time Gaussian noise of diagonal covariance $\mathbf{R}_\text{k}\equiv\mathbf{R}=4\cdot 10^{-4}\mathbf{I}$. 

\item The \emph{model} realizations consist of a main simulation (run on the same mesh used for the reference simulation) and an ensemble of $N_\text{e}=100$ coarse simulations ($8$ mesh elements per wavelength $L_0$) which are run using the numerical scheme \eqref{eq:1D_burgers_scheme_modified}. The initial condition $u(x,t=0)=u_0$ is imposed for each simulation. The outer loop provides an optimization for the value of the parameter $\theta$ driving the inlet condition. The initial condition for this parameter for each simulation is provided in the form of a Gaussian distribution $\theta \sim \mathcal{N}(0.15, \mathbf{Q}_{\theta})$, with $\mathbf{Q}_{\theta}(t=0)=6.25\cdot 10^{-4}\mathbf{I}$. 
As previously stated, the inner loop of the MGEnKF optimizes the polynomial expansion coefficients $\alpha_\text{i}$ and $\gamma_\text{i}$ controlling the behavior of the model terms introduced in the dynamical equations. Here, the physical state predicted by the main simulation is used as surrogate observation for this step. Initial values of the coefficients are described by Gaussian distributions $\alpha_\text{i} \sim \mathcal{N}(0, \mathbf{Q})$, $\gamma_\text{i} \sim \mathcal{N}(0, \mathbf{Q})$ with $\mathbf{Q}(t=0)=9\cdot 10^{-8}\mathbf{I}$ and $i=0, 1, 2, 3, 4$. The surrogate observation is here represented by the projection of the complete state predicted on the fine-grid by the main simulation over the coarse grid. 
This also implies that the operator $\mathbf{\mathcal{H}}_\text{k}^{\text{o}}\equiv\mathbf{\mathcal{H}}^{\text{o}}$ used in the inner loop of the Kalman Filter and the multigrid projector operator are the same. However, the surrogate observation sampled from the fine-grid is further randomized using a constant in time Gaussian noise of covariance $\mathbf{R}^\text{o}_\text{k}\equiv\mathbf{R}^\text{o}=1\cdot 10^{-8}\mathbf{I}$ in order to improve the convergence of the inner loop optimization procedure. As in Sec. \ref{sec:advection}, one can see that $\mathbf{R}^\text{o}$ is orders of magnitude smaller than the observation covariance matrix $\mathbf{R}$. 
The inner loop only optimizes values of $\alpha_\text{i}$ and $\gamma_\text{i}$ and no update of the state estimation is here performed. Also, the inner loop is not performed at each outer loop analysis phase, but it is instead performed around twice per characteristic time, \textit{i.e.} once every ten outer loop analyses.
\end{itemize}

\begin{figure}[ht]
\includegraphics[width=1\textwidth]{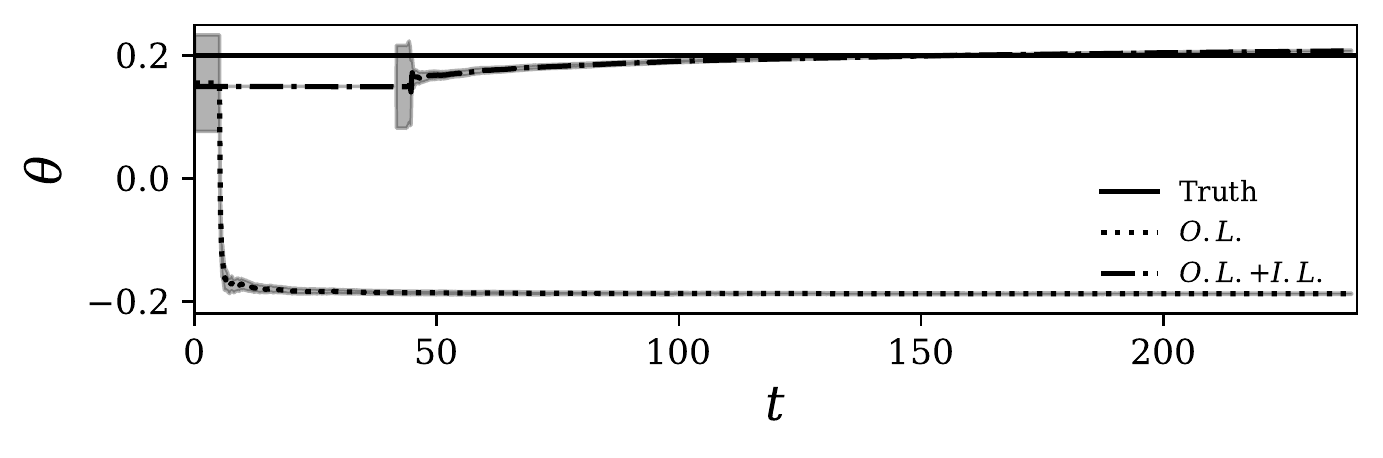}
\caption{\label{fig:amplitude_burgers_ILEL}Evolution in time of the parameter $\theta$ during the outer loop optimization via MGEnKF. Results obtained from the DA complete model (outer plus inner loop, dot-dashed line) and the simplified DA model (outer loop only, dotted line) are compared with the exact result (black line).}
\end{figure}

\begin{figure}[htbp]
\includegraphics[width=1\textwidth]{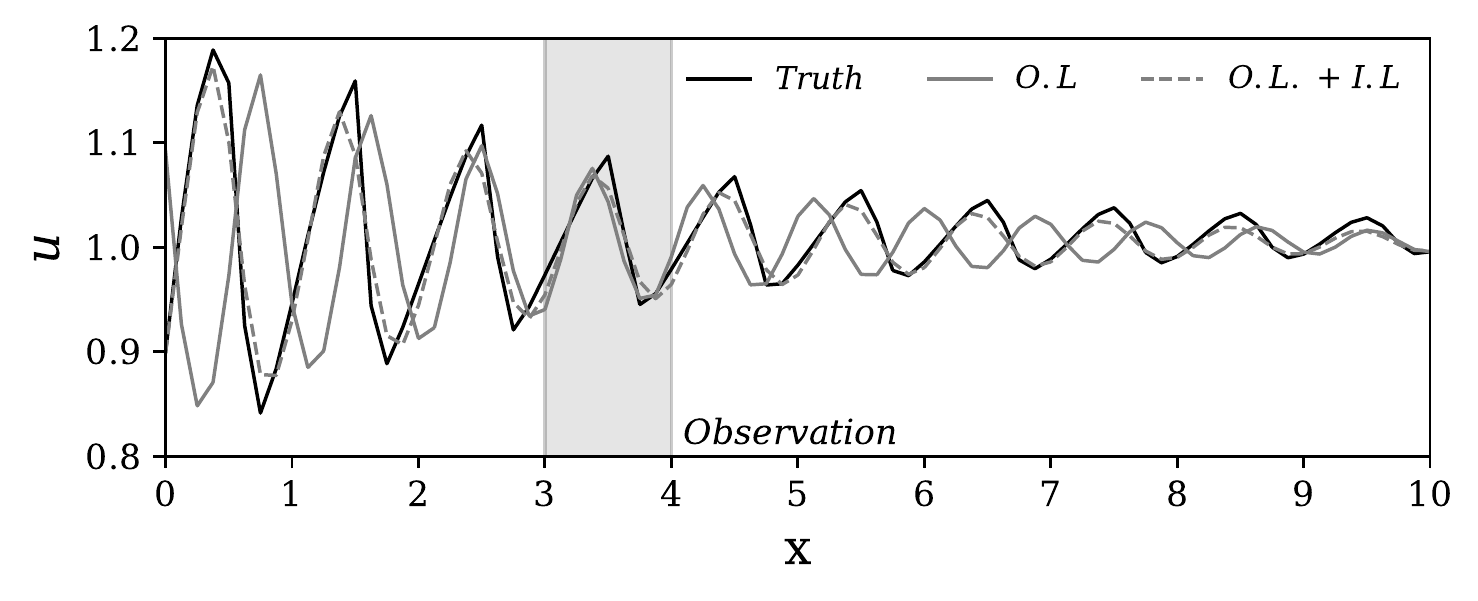}
\caption{\label{fig:burgers_CG_DA}State estimation results for the 1D Burgers' test case, projected on the coarse grid at $t=240$. The projected true state (black line) is compared with results obtained via the MGEnKF complete model (outer plus inner loop, gray dotted line) and the simplified MGEnKF model (outer loop only, gray line).}
\end{figure}

The time-evolution of the estimation for the parameter $\theta$ is shown in Fig. \ref{fig:amplitude_burgers_ILEL}. The accuracy of the complete MGEnKF scheme is remarkably good, while a significant error ($\theta=-0.2$) is observed for the simplified MGEnKF using the outer loop only. The phase error due to grid coarsening observed in Fig. \ref{fig:1D-burgers_CGmodel} for the model is responsible for this important mismatch. The reason why is clear when analysing the instantaneous physical state in Fig. \ref{fig:burgers_CG_DA}. In fact, the cumulative loss of phase of the model in the region $3\leq x \leq 4$, which includes the observation, is approximately $\pi$. Thus, this error induces a bias in the estimation of the inlet parameter $\theta$, which compensates the error in the observation region but provides massive errors outside of it. 

On the other hand, the optimization via inner loop of the coefficients $\alpha_\text{i}$ and $\gamma_\text{i}$ allows to obtain a precise representation of the flow field in the whole physical domain and not only in the observation region. The physical state obtained in Fig. \ref{fig:burgers_CG_DA} by the complete MGEnKF scheme is in better agreement with the \emph{truth}. In addition, the non-linearity of the flow is adequately captured despite the significant difference in resolution between the reference simulation and the ensemble members.  

\begin{figure}[htbp]

\includegraphics[width=1\textwidth]{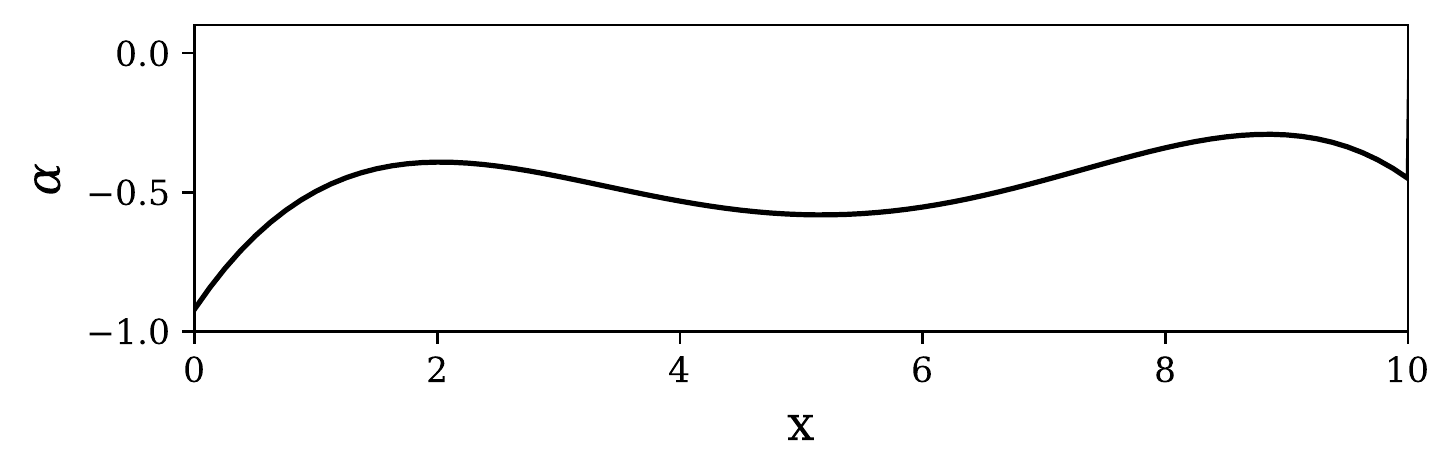}
\caption{\label{fig:alpha_burgers_ILEL}
Instantaneous space distribution of the model parameter $\alpha$ determined via inner loop optimization. The results shown correspond to a simulation time of $t=240$.}
\end{figure}

\begin{figure}[htbp]
\includegraphics[width=1\textwidth]{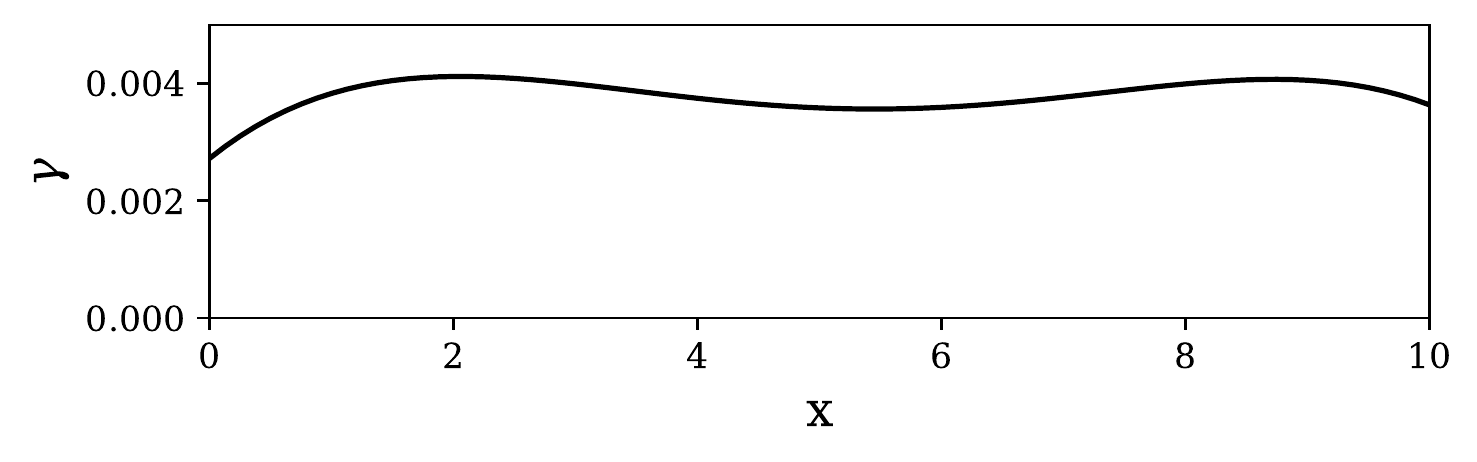}
\caption{\label{fig:gamma_burgers_ILEL}
Instantaneous space distribution of the model parameter $\gamma$ determined via inner loop optimization. The results shown correspond to a simulation time of $t=240$.}
\end{figure}
\begin{figure}[htbp]

\includegraphics[width=1\textwidth]{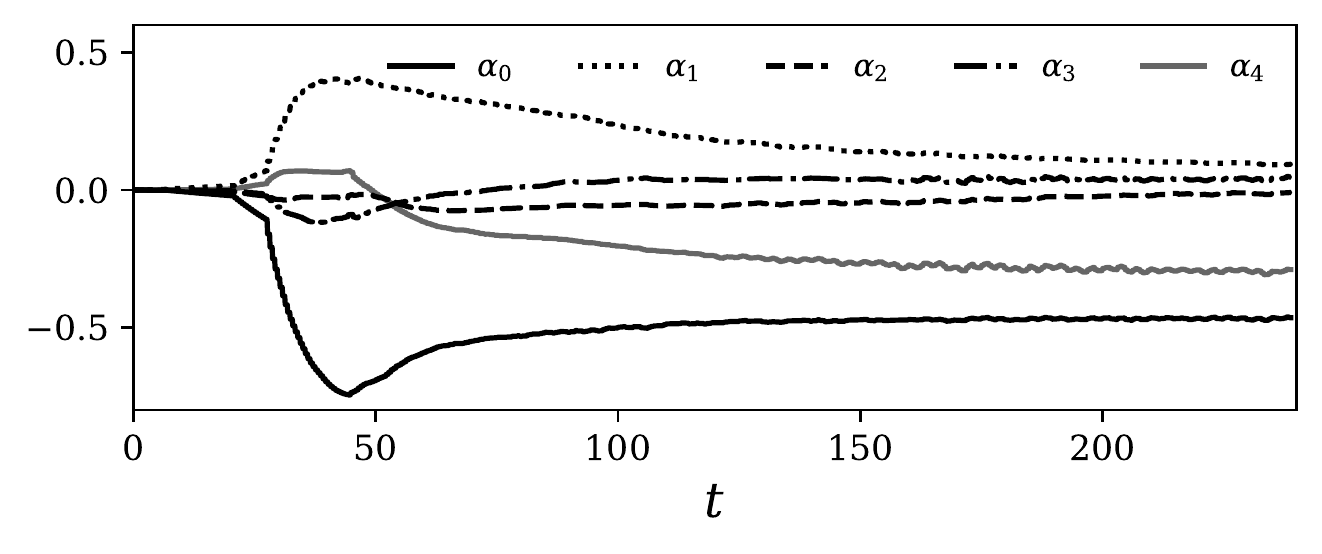}
\caption{\label{fig:calpha_burgers_ILEL}
Estimation history of the Legendre Polynomial coefficients $\alpha_\text{i}$.}
\end{figure}

\begin{figure}[htbp]
\includegraphics[width=1\textwidth]{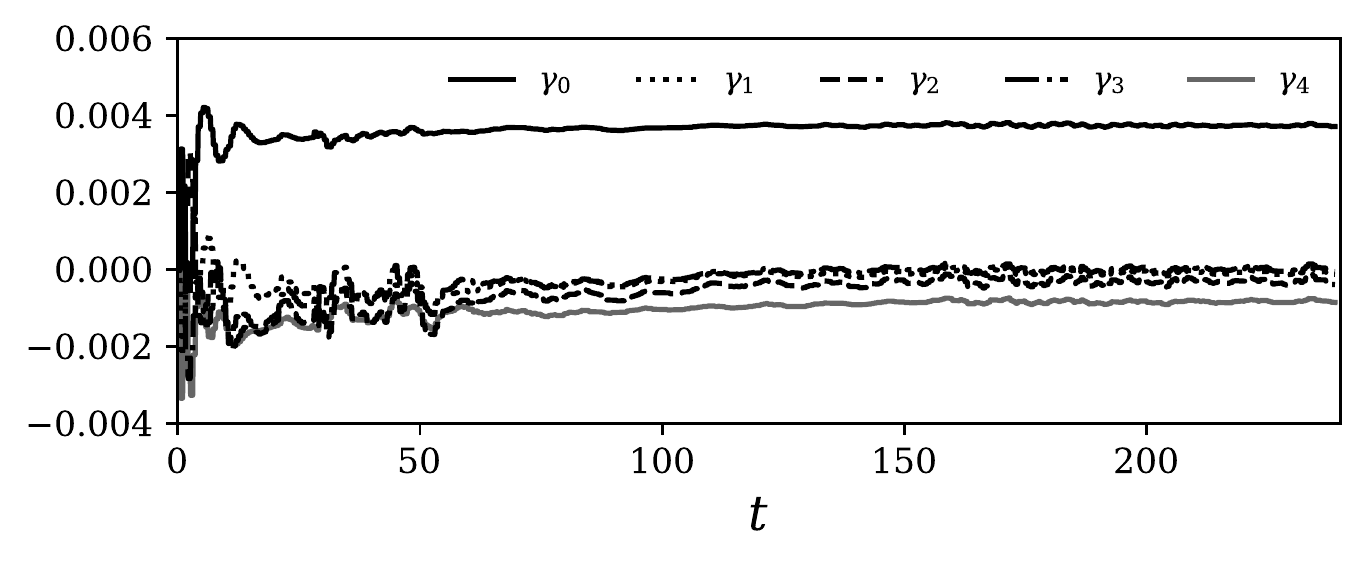}
\caption{\label{fig:cgamma_burgers_ILEL}
Estimation history of the Legendre Polynomial coefficients $\gamma_\text{i}$.}
\end{figure}

The performance of the inner loop is now investigated via the analysis of the model parameters $\alpha$ and $\gamma$. In figures \ref{fig:alpha_burgers_ILEL} and \ref{fig:gamma_burgers_ILEL}, spatial distributions of the two parameters are shown at $t=240$. It will also be shown later that time variations are weak for these quantities, so the results presented can be considered as mean values for $\alpha$ and $\gamma$ as well. As expected, $\alpha$ exhibits negative values to compensate the higher numerical diffusion due to the coarser grid. Moreover, the condition $\alpha > -1$ is strictly respected in the whole physical domain. The spatial distribution for $\gamma$ is quasi constant and equal to $\approx 0.004$. 
For the linear advection equation, if $\gamma=\sigma\left(1-\sigma^2\right)/6$, one obtains a third-order accurate scheme on the support $j-2,j-1,j,j+1$, namely the Warming, Kutler, Lomax scheme \cite{Hirsch2007}. Considering the values for $u_0$, $\Delta_x$ and $\Delta_t$ used for this analysis, it corresponds to $\gamma=0.025\left(1-0.025^2\right)/6\approx 0.004$. Thus, the optimized value for $\gamma$ obtained via the inner loop is close to the value that provides maximum accuracy in the linear advection case.

More information about the numerical models can be drawn by the analysis of the time evolution of the coefficients $\alpha_\text{i}$ and $\gamma_\text{i}$ in Fig. \ref{fig:calpha_burgers_ILEL} and \ref{fig:cgamma_burgers_ILEL}. First of all, one can see that the zero-order contributions $\alpha_\text{0}$ and $\gamma_\text{0}$ are the most important in terms of magnitude. In addition, one can also see a small but non-zero evolution in time of such coefficients. This observation is tied to the interactions between the optimization procedures performed in the inner loop and in the outer loop, whose results interact for non-linear phenomena. This is also the reason why the rate of convergence of $\alpha$, which is strictly connected to $\theta$, is slower.

Multiple strategies for the complete MGEnKF scheme have been tried varying the starting time of the inner and outer loops. It has been observed that the global optimization converges faster and it is more robust if a first phase using the inner loop only is followed by a second phase where both inner loop and outer loop are applied. This procedure allows to \emph{train} the model used for the ensemble members to perform similarly to the model employed on the fine grid for the main simulation. Therefore, the following classical DA optimization represented by the outer loop converges more rapidly to the targeted behavior provided by the observation. This initial phase of training, which has been here performed for $t \in [0, 40]$, is particularly important if the values of the parameters driving the model used for the ensemble members are unknown, like for the present analysis.

\section{Conclusions}
\label{sec:conclusions}
The predictive features of the Multigrid Ensemble Kalman Filter (MGEnKF) recently proposed for Data Assimilation of unsteady fluid flows have been investigated in this article. The analysis focused on the improvement in global performance due to the \emph{inner loop}. This step of the DA strategy targets a systematic improvement of the accuracy of the ensemble members using surrogate information from the main simulation run on the fine grid level. 

The method has been tested over two classical one-dimensional problems, namely the linear advection problem and the Burgers' equation. The results indicate the importance of the \emph{inner loop} in improving the performance of the Data Assimilation algorithm. For the linear advection case, the proposed model correction term 
$\mathbf{\mathcal{C}}_{k:k-1}$ 	
	has been derived from the exact equation in order to compensate dispersive and diffusive numerical errors. The tests performed showed that this approach can fully correct the numerical errors associated with the coarse grid level where ensemble members are run. A similar strategy has been used for the non-linear Burgers' case. Part of the correction model derived for the linear advection equation (dispersive term) has here been used as a correction term to reduce phase mismatch between fine and coarse grid forecast. An additional diffusion term parametrized by $\alpha$ was introduced to control the loss of amplitude of the solution. The improvement in the estimation accuracy due to the use of the \emph{inner loop} is remarkable. However, contrarily to what was observed in the linear advection case, the model correction term here is not able to fully correct the discrepancies due to the numerical error. This aspect, which is due to the lack of an exact correction model for the Burgers' equation,  shows the limitations of the \emph{inner loop}.

These findings open exciting perspectives of application to grid-dependent reduced-order models extensively used in fluid mechanics applications for complex flows, such as Large Eddy Simulation (LES). \textcolor{Reviewer1}{This method is extensively used in both academic and industrial studies because of its accuracy and reduced computational demands when compared with DNS. However, several research work in the literature have highlighted the extreme sensitivity of LES to variations in the set-up of the problem. Non-linear error dynamics involving different sources (discretization error, implicit / explicit filtering, subgrid scale modelling...) have been observed \cite{Meyers2006_jfm,Lucor2007_jfm,Meldi2011_pof,Meldi2012_pof,Salvetti2018_DLES} which may result in counter intuitive results such as degradation of the accuracy with mesh refinement.} An accurate model reconstruction via the \emph{inner} loop may alleviate or even prevent one of the major issues associated with multilevel applications in fluid mechanics, that is the sensitivity of the parametric description of the model (in the form of the set of parameters $\theta$) to different mesh resolution. In this scenario, one may just tune the reduced-order model for the most refined grid resolution using the \emph{outer} loop, and compensate the emerging differences from progressively coarser grids using the \emph{inner} loop to optimize the additional model. Of course, the additional model provided for the coarse grids must be suitable for this task, as seen for the one-dimensional Burgers' equation, and difficulties are expected for applications with scale-resolved turbulent flows. Combined applications of EnKF with machine learning tools, which have been recently explored for simplified test cases, may provide success in deriving precise model structures when included in the formalism of the MGEnKF model for the \emph{inner loop}.

\textbf{Acknowledgements}: Our research activities are supported by the Direction Générale de l'Armement (DGA) and the Région Nouvelle Aquitaine. 
LC would like to acknowledge the supplementary funding and excellent working conditions offered by the French Agence Nationale de la Recherche (ANR) in the framework of the projects "Closed- loop control of the wake of a road vehicle – COWAVE" (ANR-17-CE22-0008) and "Apprentissage automatique pour les récepteurs solaires à haute température – SOLAIRE" (ANR-21-CE50-0031). 

\textbf{Conflict of interest}: The Authors have no conflict of interests.

\appendix

\section{Data Assimilation algorithms}
\label{sec:DA_algorithms}
\graphicspath{{Cordier_Figs/}}
\subsection{Kalman filter algorithm}
\label{sec:KFalgo}
\begin{figure}[htbp]
\centering
\includegraphics[width=1\textwidth]{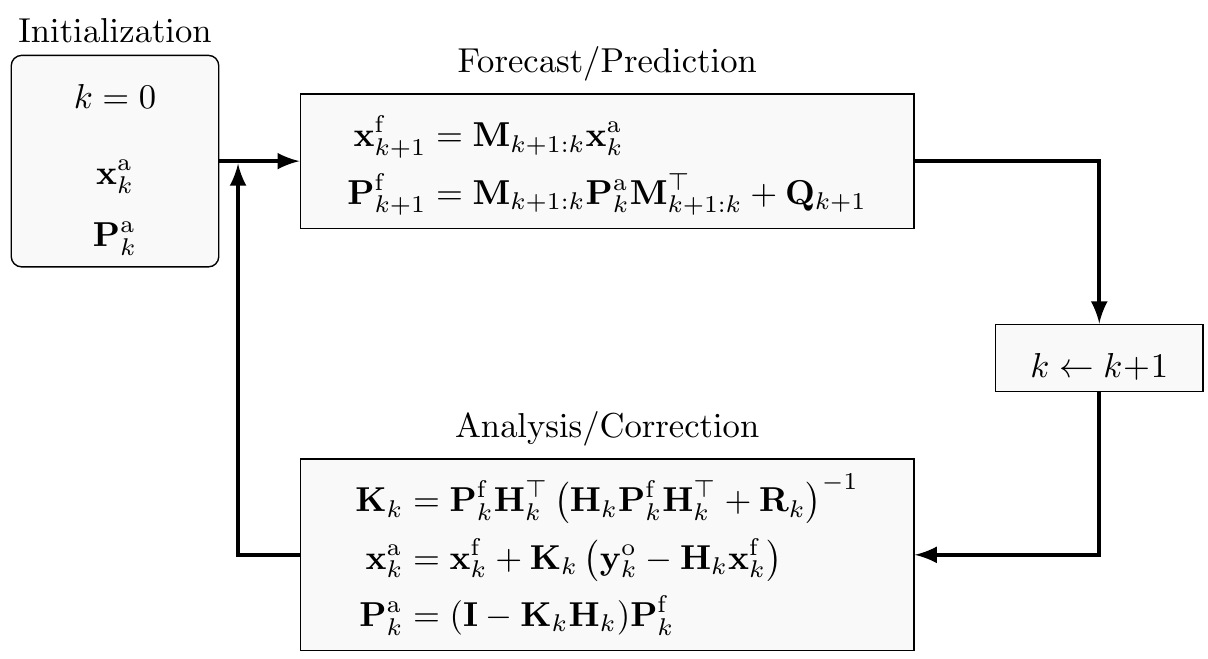}
\caption{\label{fig:KF_Init_analyse}Kalman Filter algorithm. The initialization is made with the analysed state.}
\end{figure}

The Kalman filter algorithm discussed at the beginning of Sec.~\ref{sec:maths} corresponds to Fig.~\ref{fig:KF_Init_analyse}.

\subsection{Ensemble Kalman filter algorithm}
\label{sec:EnKFalgo}
An efficient implementation of the EnKF relying on anomaly matrices is given in Algo.~\ref{alg:Stoch_EnKF_Anomaly_Carrassi}. We have used the secant method described in \cite{asch_data_2016} to change the definition of the variable $\mathbf{Y}_k^\text{f}$.

\newpage
\begin{algorithm}
\DontPrintSemicolon
\KwIn{For $k=0,\dots,K$: 
the forward models $\mathbf{\mathcal{M}}_{k:k-1}$, 
the observation models $\mathbf{\mathcal{H}}_k$, 
the observation error covariance matrices $\mathbf{R}_k$}
\KwOut{$\{\mathbf{x}_{k}^{\text{a},(i)}\}$\,;\, $k=0,\cdots,K$\,;\, $i=1,\cdots,N_\text{e}$}
\Begin{
\nlset{1:}Initialize the ensemble of forecasts $\{\mathbf{x}_{0}^{\text{f},(i)}\}$\,;\, $i=1,\cdots,N_\text{e}$\;
\For{$k=0,\dots,K$}{
\nlset{2:}Draw a statistically consistent observation set\,;\, $i=1,\cdots,N_\text{e}$\\
$\mathbf{y}_k^{\text{o},(i)}=\mathbf{y}_k^\text{o} + \mathbf{\epsilon}_k^{\text{o},(i)}\quad\text{with}\quad\mathbf{\epsilon}_k^{\text{o},(i)}\sim \mathcal{N}(0,\mathbf{R}_k)$\;
\nlset{3:}Compute the model counterparts of the observation set\,;\, $i=1,\cdots,N_\text{e}$\\
$\mathbf{y}_{k}^{\text{f},(i)}=\mathbf{\mathcal{H}}_k\left(\mathbf{x}_{k}^{\text{f},(i)}\right)$
\;
\nlset{4:}Compute the ensemble means\\
$\displaystyle\overline{\mathbf{x}_{k}^{\text{f}}}=\frac{1}{N_\text{e}}\sum_{i=1}^{N_\text{e}}\mathbf{x}_{k}^{\text{f},(i)}$\,;\,
$\displaystyle\overline{\mathbf{y}_{k}^{\text{f}}}=\frac{1}{N_\text{e}}\sum_{i=1}^{N_\text{e}}\mathbf{y}_{k}^{\text{f},(i)}$\,;\,
$\displaystyle\overline{\mathbf{\epsilon}_{k}^{\text{o}}}=\frac{1}{N_\text{e}}\sum_{i=1}^{N_\text{e}}\mathbf{\epsilon}_k^{\text{o},(i)}$
\;
\nlset{5:}Compute the normalized anomalies\,;\, $i=1,\cdots,N_\text{e}$\\
$$
\left[\mathbf{X}_k^\text{f}\right]_{:,i}=
\frac{\mathbf{x}_{k}^{\text{f},(i)}-\overline{\mathbf{x}_{k}^{\text{f}}}}{\sqrt{N_\text{e}-1}}
\,;\,
\left[\mathbf{Y}_k^\text{f}\right]_{:,i}=
\frac{%
\mathbf{y}_{k}^{\text{f},(i)}-
\overline{\mathbf{y}_{k}^{\text{f}}}
}{\sqrt{N_\text{e}-1}}
\,;\,
\left[\mathbf{E}_k^\text{o}\right]_{:,i}=
\frac{%
\mathbf{\epsilon}_k^{\text{o},(i)}-\overline{\mathbf{\epsilon}_{k}^{\text{o}}}
}{\sqrt{N_\text{e}-1}}
$$
\;
\nlset{6:}Compute the Kalman gain\\
$\mathbf{K}_k^\text{e}=
\mathbf{X}_k^\text{f}
\left(\mathbf{Y}_k^\text{f}\right)^\top
\left(
\mathbf{Y}_k^\text{f}
\left(\mathbf{Y}_k^\text{f}\right)^\top
+
\mathbf{E}_k^\text{o} \left(\mathbf{E}_k^\text{o}\right)^\top
\right)^{-1}$
\;
\nlset{7:}Update the ensemble\,;\, $i=1,\cdots,N_\text{e}$\\
$
\mathbf{x}_{k}^{\text{a},(i)}=
\mathbf{x}_{k}^{\text{f},(i)}+
\mathbf{K}_k^\text{e}
\left(
\mathbf{y}_k^{\text{o},(i)}-\mathbf{y}_{k}^{\text{f},(i)}
\right)
$
\;
\nlset{8:}Compute the ensemble forecast\,;\, $i=1,\cdots,N_\text{e}$\\
$\mathbf{x}_{k+1}^{\text{f},(i)}=\mathbf{\mathcal{M}}_{k+1:k}(\mathbf{x}_{k}^{\text{a},(i)})$
\;

}
}
\caption{\label{alg:Stoch_EnKF_Anomaly_Carrassi}Stochastic Ensemble Kalman Filter (slightly adapted from \cite{asch_data_2016}). Use of anomaly matrices with $\mathbf{Y}_k^\text{f}=\mathbf{H}_k\mathbf{X}_k^\text{f}$.}
\end{algorithm}

\subsection{Dual Ensemble Kalman filter algorithm}
\label{sec:DualEnKFalgo}
An efficient implementation of the Dual EnKF relying on anomaly matrices is given in Algo.~\ref{alg:Stoch_Dual_EnKF_Moradkhani_Carrassi}. We have slightly adapted this algorithm from \cite{DENKF_MORADKHANI2005}. 

\begin{algorithm}
\DontPrintSemicolon
\KwIn{For $k=1,\dots,K$: 
the forward models $\mathbf{\mathcal{M}}_{k:k-1}$, 
the observation models $\mathbf{\mathcal{H}}_k$, 
the observation error covariance matrices $\mathbf{R}_k$}
\KwOut{$\{\mathbf{\theta}_{k}^{\text{a},(i)}\}$ and $\{\mathbf{x}_{k}^{\text{a},(i)}\}$\,;\, $k=0,\cdots,K$}
\Begin{
\nlset{1:}Initialize $\{\mathbf{\theta}_{0}^{\text{a},(i)}\}$ and $\{\mathbf{x}_{0}^{\text{a},(i)}\}$\;
\For{$k=1,\dots,K$}{
\nlset{2:}Observation ensemble:
\vspace*{-4mm}
\begin{align*}
\mathbf{y}_k^{\text{o},(i)} & =\mathbf{y}_k^\text{o} + \mathbf{\epsilon}_k^{\text{o},(i)}\quad\text{with}\quad\mathbf{\epsilon}_k^{\text{o},(i)}\sim \mathcal{N}(0,\mathbf{R}_k)\\
\displaystyle
\mathbf{R}_k^\text{e}
& =
\frac{1}{N_\text{e}-1}
\sum_{i=1}^{N_\text{e}}
\mathbf{\epsilon}_k^{\text{o},(i)}
\left(
\mathbf{\epsilon}_k^{\text{o},(i)}
\right)^\top
\end{align*}
\;
\vspace*{-9mm}
\nlset{3:}Parameter forecast:
\vspace*{-4mm}
\begin{align*}
\mathbf{\theta}_{k}^{\text{f},(i)} & = \mathbf{\theta}_{k-1}^{\text{a},(i)}+\mathbf{\tau}_{k}^{(i)}\quad\text{with}\quad\mathbf{\tau}_{k}^{(i)}\sim \mathcal{N}(0,\mathbf{\Sigma}_k^\mathbf{\theta})\\
\mathbf{x}_{k}^{\text{f},(i)} & = \mathbf{\mathcal{M}}_{k:k-1}(\mathbf{x}_{k-1}^{\text{a},(i)},\mathbf{\theta}_{k}^{\text{f},(i)})\\
\mathbf{y}_{k}^{\text{f},(i)} & = \mathbf{\mathcal{H}}_k\left(\mathbf{x}_{k}^{\text{f},(i)}\right)
\end{align*}
\;
\vspace*{-9mm}
\nlset{4:}Compute the normalized anomalies
\vspace*{-4mm}
$$
\left[\mathbf{\Theta}_k^\text{f}\right]_{:,i}=
\frac{\mathbf{\theta}_{k}^{\text{f},(i)}-\overline{\mathbf{\theta}_{k}^{\text{f}}}}{\sqrt{N_\text{e}-1}}
\: 
;
\:
\left[\mathbf{Y}_k^\text{f}\right]_{:,i}=
\frac{%
\mathbf{y}_{k}^{\text{f},(i)}-
\overline{\mathbf{y}_{k}^{\text{f}}}
}{\sqrt{N_\text{e}-1}}
\: 
;
\:
\left[\mathbf{E}_k^\text{o}\right]_{:,i}=
\frac{%
\mathbf{\epsilon}_k^{\text{o},(i)}-\overline{\mathbf{\epsilon}_{k}^{\text{o}}}
}{\sqrt{N_\text{e}-1}}
$$
\;
\vspace*{-9mm}
\nlset{5:}Parameter update:
\vspace*{-4mm}
\begin{align*}
\mathbf{K}_k^{\theta,\text{e}} & = 
\mathbf{\Theta}_k^\text{f}
\left(\mathbf{Y}_k^\text{f}\right)^\top
\left(
\mathbf{Y}_k^\text{f}
\left(\mathbf{Y}_k^\text{f}\right)^\top
+
\mathbf{E}_k^\text{o} \left(\mathbf{E}_k^\text{o}\right)^\top
\right)^{-1}\\
\mathbf{\theta}_{k}^{\text{a},(i)} & =
\mathbf{\theta}_{k}^{\text{f},(i)}+
\mathbf{K}_k^{\theta,\text{e}}
\left(
\mathbf{y}_k^{\text{o},(i)}-\mathbf{y}_{k}^{\text{f},(i)}
\right)
\end{align*}
\;
\vspace*{-9mm}
\nlset{6:}State forecast:
\vspace*{-4mm}
\begin{align*}
\mathbf{x}_{k}^{\text{f},(i)} & = \mathbf{\mathcal{M}}_{k:k-1}(\mathbf{x}_{k-1}^{\text{a},(i)},\mathbf{\theta}_{k}^{\text{a},(i)})\\
\mathbf{y}_{k}^{\text{f},(i)} & = \mathbf{\mathcal{H}}_k\left(\mathbf{x}_{k}^{\text{f},(i)}\right)
\end{align*}
\;
\vspace*{-9mm}
\nlset{7:}Compute the normalized anomalies
\vspace*{-4mm}
$$
\left[\mathbf{X}_k^\text{f}\right]_{:,i}=
\frac{\mathbf{x}_{k}^{\text{f},(i)}-\overline{\mathbf{x}_{k}^{\text{f}}}}{\sqrt{N_\text{e}-1}}
\: 
;
\:
\left[\mathbf{Y}_k^\text{f}\right]_{:,i}=
\frac{%
\mathbf{y}_{k}^{\text{f},(i)}-
\overline{\mathbf{y}_{k}^{\text{f}}}
}{\sqrt{N_\text{e}-1}}
\: 
;
\:
\left[\mathbf{E}_k^\text{o}\right]_{:,i}=
\frac{%
\mathbf{\epsilon}_k^{\text{o},(i)}-\overline{\mathbf{\epsilon}_{k}^{\text{o}}}
}{\sqrt{N_\text{e}-1}}
$$
\;
\vspace*{-9mm}
\nlset{8:}State update:
\vspace*{-4mm}
\begin{align*}
\mathbf{K}_k^{x,\text{e}} & = 
\mathbf{X}_k^\text{f}
\left(\mathbf{Y}_k^\text{f}\right)^\top
\left(
\mathbf{Y}_k^\text{f}
\left(\mathbf{Y}_k^\text{f}\right)^\top
+
\mathbf{E}_k^\text{o} \left(\mathbf{E}_k^\text{o}\right)^\top
\right)^{-1}\\
\mathbf{x}_{k}^{\text{a},(i)} & =
\mathbf{x}_{k}^{\text{f},(i)}+
\mathbf{K}_k^{x,\text{e}}
\left(
\mathbf{y}_k^{\text{o},(i)}-\mathbf{y}_{k}^{\text{f},(i)}
\right)
\end{align*}
\;
\vspace*{-9mm}
}
}
\caption{\label{alg:Stoch_Dual_EnKF_Moradkhani_Carrassi}Dual Ensemble Kalman Filter (slightly adapted from \cite{DENKF_MORADKHANI2005}). Use of anomaly matrices with $\mathbf{Y}_k^\text{f}=\mathbf{H}_k\mathbf{X}_k^\text{f}$. We have $i=1,\cdots,N_\text{e}$.}
\end{algorithm}

\subsection{Multigrid Ensemble Kalman filter algorithm}
\label{sec:MEnKF}

\begin{algorithm}
\DontPrintSemicolon
\Begin{
\nlset{1:}Initialize 
$\left\{
\left(\mathbf{x}_{0}^\text{\tiny F}\right)^\text{a},
\overline{\bm{\theta}_{0}^\text{a}},
\bm{\theta}_{0}^{\text{a},(i)}, \overline{\bm{\psi}_{0}^\text{a}},
\bm{\psi}_{0}^{\text{a},(i)},
\left(\mathbf{x}_{0}^\text{\tiny C}\right)^{\text{a},(i)}
\right\}$\;
\For{$k=1,\dots,K$}{
\nlset{2:}Fine grid forecast:
	$$\left(\mathbf{x}_k^\text{\tiny F}\right)^\text{f}=
	\mathbf{\mathcal{M}}^\text{\tiny F}_{k:k-1}\left(
	\left(\mathbf{x}_{k-1}^\text{\tiny F}\right)^\text{a},
	\overline{\bm{\theta}_{k}^\text{a}}\right)$$
\nlset{3:}Apply \textbf{Inner/Outer loop} on the coarse grid (Algo.~\ref{alg:O_I_loop}).\\
\If {Observation available}{
\nlset{4:}Projection on the coarse grid 
$$
\left(\mathbf{x}^\text{\tiny C}_k\right)^{*}=
\Pi_\text{\tiny C}\left(\left(\mathbf{x}_k^\text{\tiny F}\right)^\text{f}\right)
$$
\nlset{5:}Fine grid state correction using the ensemble statistics:\\
		\begin{align*}
\left(\mathbf{x}^\text{\tiny C}_k\right)^{'} & =
\left(\mathbf{x}^\text{\tiny C}_k\right)^{*}+
\left(\mathbf{K}_k^\text{\tiny C}\right)^{x,\text{e}}
\left[
\left(\mathbf{y}_k^\text{\tiny C}\right)^\text{o}-
\mathbf{\mathcal{H}}_k^\text{\tiny C}
\left(\left(\mathbf{x}_k^\text{\tiny C}\right)^{*}\right)
\right]\\
\left(\mathbf{x}^\text{\tiny F}_k\right)^{'} & =
\left(\mathbf{x}^\text{\tiny F}_k\right)^\text{f}+
\Pi_\text{\tiny F}
\left(\left(
\mathbf{x}^\text{\tiny C}_k\right)^{'}-
\left(\mathbf{x}^\text{\tiny C}_k\right)^{*}\right)	
		\end{align*}
\nlset{6:} $\left(\mathbf{x}^\text{\tiny F}_k\right)^\text{a}$ is obtained through a matrix-splitting iterative procedure starting from $\left(\mathbf{x}^\text{\tiny F}_k\right)^{'}$. 
			} 
} 
} 
   \caption{Multigrid EnKF algorithm. We have $i=1,\cdots,N_\text{e}$.}
   \label{alg:algo_KF_corrected_corrected}
\end{algorithm}
%
\begin{algorithm}
\setstretch{0.9}
\KwIn{For $k=1,\dots,K$: $\mathbf{\mathcal{M}}_{k:k-1}^\text{\tiny C}$, $\mathbf{\mathcal{C}}_{k:k-1}$,  $\left(\mathbf{\mathcal{H}}_k^\text{\tiny C}\right)^\text{o}$, $\left(\mathbf{R}_k^\text{\tiny C} \right)^\text{o}$
}
\vspace{0.2cm}
\KwOut{$\bm{\theta}_{k}^{\text{a},(i)}$, $\bm{\psi}_{k}^{\text{a},(i)}$, $\left(\mathbf{x}_{k}^\text{\tiny C}\right)^{\text{a},(i)}$\,;\, $k=0,\cdots,K$}
\Begin{
\nlset{1:}Initialize $\bm{\theta}_{0}^{\text{a},(i)}$, $\bm{\psi}_{0}^{\text{a},(i)}$ and $\left(\mathbf{x}_{0}^\text{\tiny C}\right)^{\text{a},(i)}$\;
\For{$k=1,\dots,K$}{
\nlset{2:} Forecast:
\vspace*{-3mm}
\begin{align*}
\bm{\theta}_{k}^{\text{f},(i)} & = \bm{\theta}_{k-1}^{\text{a},(i)}+\bm{\tau}_{k}^{\theta,(i)}\quad\text{with}\quad\bm{\tau}_{k}^{\theta,(i)}\sim \mathcal{N}(0,\mathbf{\Sigma}_k^\mathbf{\theta})\\
\left(\mathbf{x}_{k}^\text{\tiny C}\right)^{\text{f},(i)} & = \mathbf{\mathcal{M}}_{k:k-1}^\text{\tiny C}\left(\left(\mathbf{x}_{k-1}^\text{\tiny C}\right)^{\text{a},(i)},\bm{\theta}_{k}^{\text{f},(i)}\right)+\mathbf{\mathcal{C}}_{k:k-1}\left(\left(\mathbf{x}_{k-1}^\text{\tiny C}\right)^{\text{a},(i)},\bm{\psi}_{k-1}^{\text{a},(i)}\right)
\end{align*}
Apply Inner loop (Algo.~\ref{alg:inner_loop})\\
\vspace{0.5cm}
     \If {Observation available}{
     	\vspace*{-3mm}
        \begin{align*}
		\left(\mathbf{y}_{k}^\text{\tiny C}\right)^{\text{f},(i)} & = \left(\mathbf{\mathcal{H}}_k^\text{\tiny C}\right)^\text{o}\left(\left(\mathbf{x}_{k}^\text{\tiny C}\right)^{\text{f},(i)}\right)     
		\end{align*}
		\vspace*{-1mm}
		\nlset{3:}Observation ensemble:
		\vspace*{-3mm}
		\begin{equation*}
		\left(\mathbf{y}_k^\text{\tiny C}\right)^{\text{o},(i)}  =\left(\mathbf{y}_k^\text{\tiny C}\right)^\text{o} + \left(\bm{\epsilon}_k^\text{\tiny C}\right)^{\text{o},(i)}\quad\text{with}\quad\left(\bm{\epsilon}_k^\text{\tiny C}\right)^{\text{o},(i)}\sim \mathcal{N}(0,\left(\mathbf{R}_k^\text{\tiny C} \right)^\text{o})
		\end{equation*}
		\vspace*{-3mm}
		\nlset{4:}Compute the normalized anomalies
		\vspace*{-0mm}
		$$
		\left[\mathbf{\Theta}_k^\text{f}\right]_{:,i}=
		\frac{\bm{\theta}_{k}^{\text{f},(i)}-\overline{\bm{\theta}_{k}^{\text{f}}}}{\sqrt{N_\text{e}-1}}
		\: 
		;
		\:
		\left[\mathbf{Y}_k^\text{f}\right]_{:,i}=
		\frac{%
		\left(\mathbf{y}_{k}^\text{\tiny C}\right)^{\text{f},(i)}-
		\overline{\left(\mathbf{y}_{k}^\text{\tiny C}\right)^{\text{f}}}
		}{\sqrt{N_\text{e}-1}}
		$$
		\vspace*{-3mm}
		\nlset{5:}Parameter update:
		\vspace*{-1mm}
		\begin{align*}
		\left(\mathbf{K}_k^\text{\tiny C}\right)^{\theta,\text{e}} & = 
		\mathbf{\Theta}_k^\text{f}
		\left(\mathbf{Y}_k^\text{f}\right)^\top
		\left(
		\mathbf{Y}_k^\text{f}
		\left(\mathbf{Y}_k^\text{f}\right)^\top
		+
		\left(\mathbf{R}_k^\text{\tiny C} \right)^\text{o}
		\right)^{-1}\\
		\bm{\theta}_{k}^{\text{a},(i)} & =
		\bm{\theta}_{k}^{\text{f},(i)}+
		\left(\mathbf{K}_k^\text{\tiny C}\right)^{\theta,\text{e}}
		\left(
		\left(\mathbf{y}_k^\text{\tiny C}\right)^{\text{o},(i)}-\left(\mathbf{y}_{k}^\text{\tiny C}\right)^{\text{f},(i)}
		\right)
		\end{align*}
		\nlset{6:}State re-forecast (only if pure dual formulation is retained, otherwise the state update can be evaluated with the first forecast $\left(\mathbf{x}_{k}^\text{\tiny C}\right)^{\text{f},(i)}$) :
		\vspace*{-3mm}
		\begin{align*}
		\left(\mathbf{x}_{k}^\text{\tiny C}\right)^{\text{f},(i)} & = \mathbf{\mathcal{M}}_{k:k-1}^\text{\tiny C}\left(\left(\mathbf{x}_{k-1}^\text{\tiny C}\right)^{\text{a},(i)},\bm{\theta}_{k}^{\text{a},(i)}\right)+\mathbf{\mathcal{C}}_{k:k-1}\left(\left(\mathbf{x}_{k-1}^\text{\tiny C}\right)^{\text{a},(i)},\bm{\psi}_{k-1}^{\text{a},(i)}\right)\\
		\left(\mathbf{y}_{k}^\text{\tiny C}\right)^{\text{f},(i)} & = \left(\mathbf{\mathcal{H}}_k^\text{\tiny C}\right)^\text{o}\left(\left(\mathbf{x}_{k}^\text{\tiny C}\right)^{\text{f},(i)}\right)
		\end{align*}
		\vspace*{-3mm}
		\nlset{7:}Compute the normalized anomalies
		\vspace*{-1mm}
		$$
		\left[\mathbf{X}_k^\text{f}\right]_{:,i}=
		\frac{\left(\mathbf{x}_{k}^\text{\tiny C}\right)^{\text{f},(i)}-\overline{\left(\mathbf{x}_{k}^\text{\tiny C}\right)^{\text{f}}}}{\sqrt{N_\text{e}-1}}
		\: 
		;
		\:
		\left[\mathbf{Y}_k^\text{f}\right]_{:,i}=
		\frac{%
		\left(\mathbf{y}_{k}^\text{\tiny C}\right)^{\text{f},(i)}-
		\overline{\left(\mathbf{y}_{k}^\text{\tiny C}\right)^{\text{f}}}
		}{\sqrt{N_\text{e}-1}}
		$$
		\vspace*{-5mm}
		\nlset{8:}State update:
		\vspace*{-0mm}
		\begin{align*}
		\left(\mathbf{K}_k^\text{\tiny C}\right)^{x,\text{e}} & = 
		\mathbf{X}_k^\text{f}
		\left(\mathbf{Y}_k^\text{f}\right)^\top
		\left(
		\mathbf{Y}_k^\text{f}
		\left(\mathbf{Y}_k^\text{f}\right)^\top
		+
		\left(\mathbf{R}_k^\text{\tiny C} \right)^\text{o}
		\right)^{-1}\\
		\left(\mathbf{x}_{k}^\text{\tiny C}\right)^{\text{a},(i)} & =
		\left(\mathbf{x}_{k}^\text{\tiny C}\right)^{\text{f},(i)}+
		\left(\mathbf{K}_k^\text{\tiny C}\right)^{x,\text{e}}
		\left(
		\left(\mathbf{y}_k^\text{\tiny C}\right)^{\text{o},(i)}-\left(\mathbf{y}_{k}^\text{\tiny C}\right)^{\text{f},(i)}
		\right)
		\end{align*}
     } 
} 
} 
   \caption{Inner/Outer loop of the MGEnKF applied on the coarse grid. Inspired from Algo.~\ref{alg:Stoch_Dual_EnKF_Moradkhani_Carrassi}. We have $i=1,\cdots,N_\text{e}$.}
   \label{alg:O_I_loop}
\end{algorithm}
%
\begin{algorithm}
\setstretch{0.9}
\KwIn{For $k=1,\dots,K$: 
the ensemble coarse grid forecast $\left(\mathbf{x}_{k}^\text{\tiny C}\right)^{\text{f},(i)}$, 
the surrogate observation operator $\left(\mathbf{\mathcal{H}}_k^\text{\tiny C}\right)^\text{so}$, 
the surrogate observation error covariance matrix $\left(\mathbf{R}_k^\text{\tiny C}\right)^\text{so}$}
\KwOut{$\bm{\psi}_{k}^\text{\text{a},(i)}$\,;\, $k=0,\cdots,K$}
\Begin{
\nlset{1:}Initialize $\bm{\psi}_{0}^\text{\text{a},(i)}$\;
\For{$k=1,\dots,K$}{
\nlset{2:}Parameter forecast:
\vspace*{-3mm}
\begin{equation*}
\bm{\psi}_{k}^{\text{f},(i)}  = \bm{\psi}_{k-1}^{\text{a},(i)}+\mathbf{\tau}_{k}^{\psi,(i)}\quad\text{with}\quad\mathbf{\tau}_{k}^{\psi,(i)}\sim \mathcal{N}(0,\mathbf{\Sigma}_k^\mathbf{\psi})
\end{equation*}
     \If {Surrogate observation available}{
     	\vspace*{-3mm}
        \begin{align*}
		\left(\mathbf{y}_{k}^\text{\tiny C}\right)^{\text{f},(i)} & = \left(\mathbf{\mathcal{H}}_k^\text{\tiny C}\right)^\text{so}\left(\left(\mathbf{x}_{k}^\text{\tiny C}\right)^{\text{f},(i)}\right)     
		\end{align*}
		\vspace*{-1mm}
		\nlset{3:}Observation ensemble:
		\vspace*{-3mm}
		\begin{equation*}
		\left(\mathbf{y}_k^\text{\tiny C}\right)^{\text{so},(i)}  =\left(\mathbf{y}_k^\text{\tiny C}\right)^\text{so} + \left(\mathbf{\epsilon}_k^\text{\tiny C}\right)^{\text{so},(i)}\quad\text{with}\quad\left(\mathbf{\epsilon}_k^\text{\tiny C}\right)^{\text{so},(i)}\sim \mathcal{N}(0,\left(\mathbf{R}_k^\text{\tiny C}\right)^\text{so})
		\end{equation*}
		\vspace*{-3mm}
		\nlset{4:}Compute the normalized anomalies
		\vspace*{-0mm}
		$$
		\left[\mathbf{\Psi}_k^\text{f}\right]_{:,i}=
		\frac{\bm{\psi}_{k}^{\text{f},(i)}-\overline{\bm{\psi}_{k}^{\text{f}}}}{\sqrt{N_\text{e}-1}}
		\: 
		;
		\:
		\left[\mathbf{Y}_k^\text{f}\right]_{:,i}=
		\frac{%
		\left(\mathbf{y}_{k}^\text{\tiny C}\right)^{\text{f},(i)}-
		\overline{\left(\mathbf{y}_{k}^\text{\tiny C}\right)^{\text{f}}}
		}{\sqrt{N_\text{e}-1}}
		$$
		\vspace*{-3mm}
		\nlset{5:}Parameter update:
		\vspace*{-1mm}
		\begin{align*}
		\left(\mathbf{K}_k^\text{\tiny C}\right)^{\psi,\text{e}} & = 
		\bm{\Psi}_k^\text{f}
		\left(\mathbf{Y}_k^\text{f}\right)^\top
		\left(
		\mathbf{Y}_k^\text{f}
		\left(\mathbf{Y}_k^\text{f}\right)^\top
		+
		\left(\mathbf{R}_k^\text{\tiny C}\right)^\text{so}
		\right)^{-1}\\
		\bm{\psi}_{k}^{\text{a},(i)} & =
		\bm{\psi}_{k}^{\text{f},(i)}+
		\left(\mathbf{K}_k^\text{\tiny C}\right)^{\psi,\text{e}}
		\left(
		\left(\mathbf{y}_k^\text{\tiny C}\right)^{\text{so},(i)}-\left(\mathbf{y}_{k}^\text{\tiny C}\right)^{\text{f},(i)}
		\right)
		\end{align*}

     } 
 }
} 
   \caption{Inner loop of the MGEnKF algorithm applied on the coarse mesh. We have $i=1,\cdots,N_\text{e}$.}
   \label{alg:inner_loop}
\end{algorithm}

Algorithm \ref{alg:algo_KF_corrected_corrected} represents a simplified, ready-to-use, application of the conceptual methodology presented in 
Sec.~\ref{sec:MGEnKF_Description} and Figs.~\ref{fig:schema_MGENKF} and \ref{fig:schema_inou_loop}.
\bibliography{Bibliography_HDR}

\end{document}